\documentclass[aps,pra,twocolumn,preprintnumbers,longbibliography,floatfix]{revtex4-2}

\usepackage{graphicx}  
\usepackage{subfigure}
\usepackage{multirow}
\usepackage{float}
\usepackage{fancyhdr}
\usepackage{longtable}
\usepackage{parskip}
\usepackage[T1]{fontenc}
\usepackage{dcolumn}   
\usepackage{amsmath}
\usepackage{hyperref}
\usepackage[dvipsnames]{xcolor}
\usepackage{soul}
\usepackage[utf8]{inputenc}
\usepackage[english]{babel}
\usepackage{bm}        
\usepackage{amsfonts}  
\usepackage{amsmath}   
\usepackage{amssymb}   
\usepackage{mathtools}
\usepackage{soul}
\usepackage{comment}

\newcommand{\pwisein}{\left\{ \begin{array}{ll}}
\newcommand{\pwiseout}{\end{array}\right.}

\begin{document}

\title{Trotter errors from dynamical structural instabilities of Floquet maps in quantum simulation}


\author{Karthik Chinni}
\email{kchinni@unm.edu}
\affiliation{Center for Quantum Information and Control, Department of Physics 
and Astronomy, University of New Mexico, Albuquerque, New Mexico 87131, USA}
\author{Manuel H. Muñoz-Arias}
\affiliation{Center for Quantum Information and Control, Department of Physics 
and Astronomy, University of New Mexico, Albuquerque, New Mexico 87131, USA}
\author{Ivan H. Deutsch}
\affiliation{Center for Quantum Information and Control, Department of Physics 
and Astronomy, University of New Mexico, Albuquerque, New Mexico 87131, USA}
\author{Pablo M. Poggi}
\affiliation{Center for Quantum Information and Control, Department of Physics 
and Astronomy, University of New Mexico, Albuquerque, New Mexico 87131, USA}

\begin{abstract}
We study the behavior of errors in the quantum simulation of spin systems with long-range multi-body interactions resulting from the Trotter-Suzuki decomposition of the time-evolution operator. We identify a regime where the Floquet operator underlying the Trotter decomposition undergoes sharp changes even for small variations in the simulation step size. This results in a time evolution operator that is very different from the dynamics generated by the targeted Hamiltonian, which leads to a proliferation of errors in the quantum simulation. These regions of sharp change in the Floquet operator, referred to as structural instability regions, appear typically at intermediate Trotter step sizes and in the weakly-interacting regime, and are thus complementary to recently revealed quantum chaotic regimes of the Trotterized evolution \cite{siberer2019,Heyl2019}. We characterize these structural instability regimes in $p$-spin models, transverse-field Ising models with all-to-all $p$-body interactions, and analytically predict their occurrence based on unitary perturbation theory. We further show that the effective Hamiltonian associated with the Trotter decomposition of the unitary time-evolution operator, when the Trotter-step size is chosen to be in the structural instability region, is very different from the target Hamiltonian, which explains the large errors that can occur in the simulation in the regions of instability. These results have implications for the reliability of  near-term gate-based quantum simulators, and reveal an important interplay between errors and the physical properties of the system being simulated.
\end{abstract}

\date{\today}

\maketitle 
\section{Introduction}
\label{sec:introduction}
A primary application of quantum computers is simulation of quantum many-body systems that are classically intractable. Quantum simulation has applications in a wide variety of fields such as quantum chemistry \cite{Georgescu2014, Aspuru2005}, high-energy physics \cite{Martinez2016, Klco2018}, condensed-matter physics \cite{Ebadi2021, Hofstetter2018} and quantum machine learning \cite{Biamonte2017, Gao2017}. The devices of the current era, characterized by intermediate system size ($\sim 100$ qubits) and lack of full fault-tolerant error correction, are referred to as noisy intermediate scale-quantum (NISQ) devices. The goal of NISQ-era devices is to perform less-demanding tasks than required for universal quantum computer, but ones that can still surpass the capability of the classical computers \cite{Preskill2018, Deutsch2020}. Recently, there have been a number of studies about achieving quantum advantage in NISQ devices in the context of quantum simulation \cite{Ebadi2021,Scholl2021,Zhang2017,Bernien2017,Bloch2012}, optimization \cite{Peruzzo2014} and sampling \cite{Arute2019}.

To implement quantum simulation in a gate-based architecture, a common approach is to approximate the time-evolution operator generated by target simulation Hamiltonian using the Trotter-Suzuki decomposition~\cite{lloyd1996,Childs2018, Childs2021, Yi2021a,Yi2021b}. Consider the target time-independent Hamiltonian, $H_{\rm tar}$, given by
\begin{equation}
\label{eq:hami_decomposition}
    H_{\text{tar}}=H_{1}+H_{2},
\end{equation} 
and suppose that the time evolution operator associated with each of the individual terms in the target Hamiltonian, $\{e^{-iH_{1}t},e^{-iH_{2}t}\}$, can be implemented. Then, the target time-evolution operator $U_{\text{tar}}=\exp{(-i H_{\text{tar}} t)}$ can be implemented through the first-order Trotter-Suzuki decomposition given by 
\begin{align}
\label{eq:trotterized_unitary}
    U_{\text{trot}} =\bigl(e^{-iH_{1}\frac{t}{n}}e^{-iH_{2}\frac{t}{n}} \bigr)^{n}\equiv (U_{\delta}(\tau))^{n},
\end{align}
with the Trotter-step size given by $\tau=t/n$. In the limit $n \rightarrow \infty$, the Trotterized unitary in Eq. (\ref{eq:trotterized_unitary}) becomes identical to $U_{\text{tar}}$. In practice, $n$ is a finite number and leads to errors in the overall simulation, which are bounded by 
\begin{equation}
\label{eq:trotterized_unitary_error}
    ||U_{\rm trot}-U_{\rm tar}||\leq \frac{t^{2}}{2n}\lVert[H_{1 },H_{2}]\rVert \; ,
\end{equation}
where $||.||$ is the spectral norm. Here, $n=\mathcal{O}\bigl(\lVert[H_{1 },H_{2}]\rVert t^{2}/\epsilon \bigr)$ is chosen so that the simulation has an overall accuracy $\epsilon$ \cite{Childs2021}. Henceforth, we will refer to 
the errors resulting from the Trotter-Suzuki decomposition as Trotter errors.

In recent works, the errors resulting from Trotter approximation have been analyzed by associating the unitary resulting from Trotter-Suzuki decomposition $U_{\rm trot}$ with the Floquet operator of a time-dependent periodically-``kicked" Hamiltonian $H_{\delta}(t)$ \cite{siberer2019,Heyl2019}. This analysis revealed the existence of a regime in which dynamics of low-order observables given by the kicked Hamiltonian yields an accurate approximation to that of the target time-independent Hamiltonian. However, this regime breaks down for large enough Trotter step size, where errors are not controlled anymore, and the Floquet dynamics significantly deviate from the target dynamics. This uncontrolled error regime was attributed to the fact that kicked system develops quantum-chaotic features that are absent in the target Hamiltonian. 

The existence of such nontrivial crossover in the Trotter error behavior reveals that the physical properties of the simulated Hamiltonian obtained via mapping to an effective time-dependent Floquet system can determine behavior of quantum-simulator errors. A related recent study, involving the simulation of adiabatic dynamics, characterized the emergence of quantum simulation errors to the closing of the spectral gap in the simulated Hamiltonian~\cite{Yi2021a,Yi2021b}. These works demonstrate that the physics associated with Floquet dynamics can affect how strongly errors saturate the Trotter-error bounds.

In this work we identify a new physical mechanism whereby the Floquet dynamics associated with Trotter-Suzuki decomposition lead to large errors in quantum simulation.
 Specifically, we study quantum simulation of $p-$spin models~\cite{Jorg2010,matsuura2017, Filippone2011, Bapst2012, Munoz2020}, which describe long-range interacting systems with multiple-body interactions, or equivalently, the dynamics of a large collective spin with nonlinear evolution to the $p^{\mathrm{th}}$ power. Such models are mean-field models, where the mean-field dynamics exactly describes the thermodynamic/classical limit~\cite{Bapst2012}. The classical dynamics on the phase space thus informs us about the potential mechanisms for errors in a quantum simulator. 
 While the target mean-field collective spin models we will study here are classically simulable, they represent useful models for benchmarking quantum simulators. That is, these models can be used to analyze the reliability of the output of a quantum simulator by comparing its output with the classically obtained result.
 
 We show that, even in the absence of chaos, there are parameter regimes where bifurcations arise in the mean-field dynamics associated with $U_{\rm trot}$ that can cause a qualitative change in the structure of the classical phase space. These bifurcations appear only in the classical limit of the Floquet map corresponding to $U_{\rm trot}$ and are absent in the ideal target unitary $U_{\text{tar}}$, signaling the existence of regimes where the target and simulated dynamics strongly diverge. These dynamical instabilities can lead to large errors in the expected value of various observables.
We name these parameter regions ``structural instabilities'' of the simulated unitary $U_{\text{trot}}$, and characterize them in detail for $p$-spin models.


As the mean-field dynamics are equivalent to the classical dynamics in the thermodynamic limit, we study the structure of the classical phase space associated with $U_{\rm trot}(\tau)$. We show that the structure of phase space undergoes significant changes accommodated by multiple bifurcations in all the instability regions as $\tau$ is varied by a small amount. A precursor of this phenomenon was previously noted in Ref. \cite{Munoz2021}, where the route from integrability to chaos was studied for periodically kicked $p$-spin systems. Here, we show that that this phenomenon does not require us to be in the thermodynamic limit, and use unitary perturbation theory to show that in the regime of structural instability, the eigenstates of the Trotterized unitary substantially deviate from those of target-unitary eigenstates, leading to an effective Hamiltonian $H_{\rm eff}$   associated with $U_{\rm trot}=e^{-i H_{\rm eff}\tau}$, that is very different from the target Hamiltonian. Crucially, our analysis shows that structural instabilities appear at Trotter step sizes which are typically smaller than those required for the system to transition to chaos, making this novel regime particularly relevant to understanding the behavior of errors in near-term quantum simulators. 


The remainder of the manuscript is organized as follows. In Sec. II, we discuss the mapping between quantum simulation via Trotter-Suzuki decomposition
and Floquet dynamics, and discuss the intuition behind the emergence of quantum simulation errors from a classical perspective. We then introduce $p$-spin models and analyze two different regions of the parameter space that result in large errors as a result of Trotterization for a general Hamiltonian: the chaotic region and structural instability region caused by bifurcations. Then we specifically analyze how the structural instability regions lead to large errors in the Trotter approximation of the time-evolution operator for $p$-spin models. 
Following this, in Sec. III, we study the regions of structural instabilities through the use of unitary perturbation theory, which allows us to analytically predict the behavior of errors in long-time averaged magnetization. In Sec. IV, we construct the effective Hamiltonian associated with a general structural instability region in the $p$-spin Hamiltonian, which explains the presence of large errors in various observables and the significant changes taking place on the classical phase space in these regions. Finally, we show that the effective Hamiltonian construction accurately describes the appearance of unstable fixed points, and in particular predicts the growth-rate of the out-of-time-order correlator (OTOC).

\section{Trotter errors in quantum simulation of $p$-spin models}
In this section we describe the connection between the Trotter-Suzuki decomposition of the target unitary evolution operator and Floquet dynamics, and discuss the implications of Trotterization in the classical limit in terms of Hamiltonian flows and area-preserving maps. We then introduce $p$-spin models and their kicked counterparts, and use these models to illustrate the different possible scenarios that could lead to the proliferation of Trotter errors.

\subsection{Trotterized evolution and kicked systems: quantum and classical}
\label{sec:sec2}
In a quantum simulator, the time evolution map generated by the desired target Hamiltonian $H_{\rm tar}$ can be implemented through the use of the Trotterized unitary given in Eq. (\ref{eq:trotterized_unitary}). This unitary map can be analyzed by identifying $U_{\delta}(\tau)$ as the time evolution operator generated by a time-dependent periodic Hamiltonian $H_\delta(t)$, which takes the the form
\begin{equation}
    H_\delta(t) = H_1 + \tau f_\tau(t)H_2,\ \mathrm{where}\ f_\tau(t)=\sum\limits_{n=-\infty}^{n=\infty} \delta(t-n\tau).
    \label{eq:kicked_hamiltonian}
\end{equation}
The corresponding unitary evolution for one time period is given by the Floquet operator,
\begin{align}
\begin{split}
    U_{\delta}(\tau)&=\mathcal{T}\bigl[\int_{0}^{\tau}dt'\exp\bigl(-iH_{\delta}(t')\bigr)\bigr] \\
&=\exp(-i\tau H_{1})\exp(-i\tau H_{2}),
\end{split}
\end{align}
where $\mathcal{T}$ is the time-ordering operator.  Due to the impulse-like driving present in Eq. (\ref{eq:kicked_hamiltonian}), these are sometimes called ``kicked'' systems. Since $H_{\text{tar}}$ and $H_{\delta}(\tau)$ are different Hamiltonians except in the limit $\tau \rightarrow 0$, the Trotterized simulation is expected to be different from the ideal Hamiltonian evolution, resulting in simulation errors for a finite-sized Trotter step size $\tau$, as shown in Eq. (\ref{eq:trotterized_unitary_error}). 

A way of studying the physical mechanisms behind these errors is to rewrite Eq. (\ref{eq:kicked_hamiltonian}) as
\begin{equation}
    H_\delta(t) = H_1 + H_2 + g_\tau(t)H_2
\label{eq:kicked_hamiltonian2}
\end{equation}
\noindent where $g_\tau(t)=\tau f_{\tau}(t)-1$. Since $H_{tar}=H_1+H_2$, the evolution of $H_\delta(t)$ corresponds to that of $H_{tar}$ under the action of an additional time-dependent perturbation, whose action is weak for small $\tau$, as can be deduced from Eq. (\ref{eq:trotterized_unitary_error}). If $H_{\mathrm{tar}}$ corresponds to an integrable Hamiltonian, the inclusion of a time-dependent perturbation is expected to break such integrability. 
The proliferation of errors in the quantum simulation of an integrable system that becomes chaotic as a result of Trotterization was discussed in \cite{siberer2019} for the case of the quantum kicked top. Furthermore, a similar behavior was found even when considering quantum Hamiltonians with many-body quantum chaos, where the classical limit is not clear cut \cite{Heyl2019}. From this picture, we can expect the  transition from regularity to chaos in the Trotterized unitary to be a fairly general phenomenon. 

In contrast, here we will focus on a regime where the perturbation is not strong enough to make the system chaotic, but nonetheless has the potential to make the perturbed dynamics significantly different from the target dynamics. In classical Hamiltonian systems,  the Kolmogorov–Arnold–Moser (KAM) theorem guarantees that for sufficiently small $\tau$, the regularity of the original Hamiltonian is preserved \cite{Reichl2004}, and thus one expects that the perturbed evolution to be still a good approximation for the ideal dynamics. However, there are situations like the presence of resonant regular orbits, which fall outside the validity of the KAM theorem \cite{Wimberger2014} and in which even small perturbations can have a big effect in the dynamics of the system. These changes are signaled by the emergence of new fixed points through bifurcations, among other mechanisms. As this happens, the emergence of instabilities and the development of significant changes in the phase space structures also have the potential to substantially modify the dynamics of the effective system $H_{\delta}$ from that of the ideal target Hamiltonian $H_{tar}$.

As we will show, for some models these features can  persist in quantum dynamics far from the classical limit, and these instabilities determine parameter regimes of high Trotter errors. Moreover, since these features appear at smaller perturbation strengths (before the transition to chaos), they would affect the Trotterized evolution at smaller values of the Trotter-step size $\tau$, making them particularly relevant for quantum simulation.
A way to identify these high Trotter-error regions in the quantum regime is to determine the regions where the eigenstates of $U_{\delta}$ are very different from $U_{\rm tar}$. This follows from the correspondence principle. The quasi-probability distribution (e.g., Husimi distribution) of the eigenstates associated with the time-evolution operator is expected to have a significant overlap with the corresponding phase-space trajectories, when both of them are plotted on the classical phase space \cite{Trail2008}.
Thus, the eigenstates of $U_{\delta}$ and $U_{\rm tar}$ will be very different whenever the corresponding classical trajectories are different. In the following we will explore this phenomenon in the quantum simulation of long-range interacting spin models, whose mean-field limit is equivalent to the thermodynamic/classical limit, by analyzing the eigenstates of $U_{\delta}$ and $U_{\rm tar}$. In addition, as we will discuss, some of the underlying physical mechanisms behind these phenomena are much more general, and are expected to appear in other types of quantum many-body systems. 
\begin{figure}[t]
\centering
\includegraphics[width=\columnwidth]{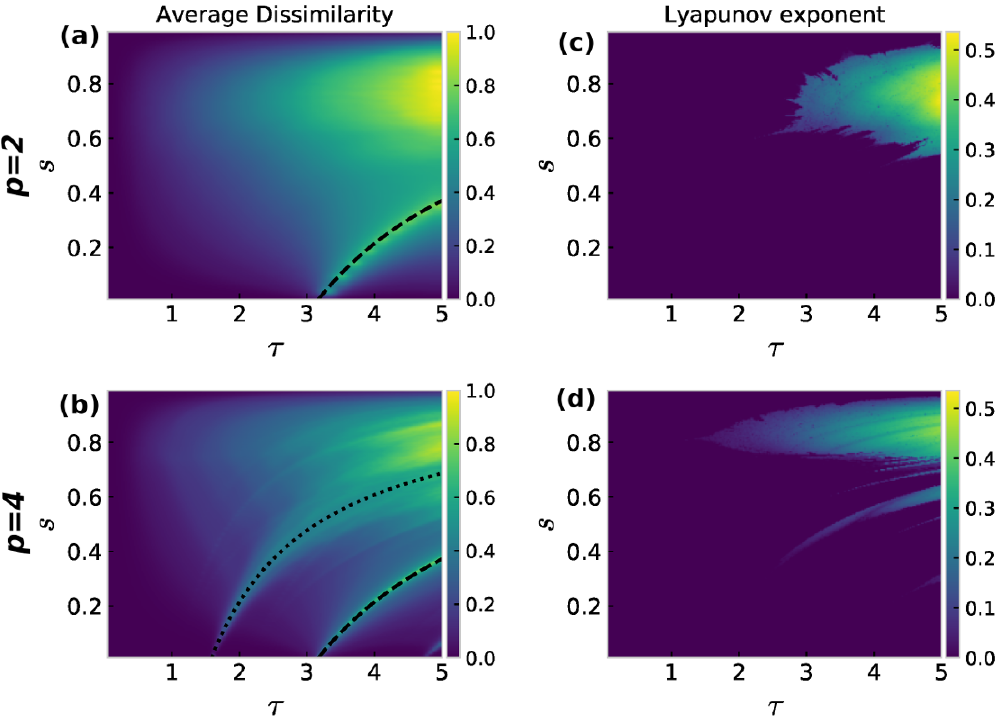}
\caption{\textbf{(a,b)} Average dissimilarity of the Floquet eigenstates of $H_{\delta}(\tau)$ with respect to the eigenstates of the $p$-spin Hamiltonian, Eq. (\ref{eq:dissimilarity_definition}).  \textbf{(c,d)} Largest Lyapunov exponent of the classical stroboscopic map associated with the Floquet map of the simulator as a function of $(\tau, s)$. In both cases we show the results for $p = 2,4$, top and bottom, respectively, and $N = 128$. The dashed and dotted lines in \textbf{(a,b)}  have the form $s(\tau) = 1 - \frac{\alpha}{\tau}$, with $\alpha = \pi, \pi/2$, respectively. These represent regions of structural instability that are not associated with chaos as the largest Lyapunov exponent corresponding to these regions is zero.}
\label{fig:fig1}
\end{figure}
\subsection{$p$-spin models}
\label{sec:p_spin_generalities}
The family of magnetic models usually referred to as $p$-spin models describes a collection of $N$ spin-$1/2$ particles on a fully connected graph, interacting through $p$-body Ising-like coupling in the presence of an external transverse magnetic field. This collection of particles experience two different orderings, a paramagnetic phase induced by an external homogeneous magnetic field, and a ferromagnetic phase induced by a $p$-body Ising-like interaction. The Hamiltonian for this family of models is given by 
\begin{align} \label{eqn:p_spin_hami}
    \begin{split}
            H&=-\frac{h}{2}\sum_{i=1}^{N} \sigma_{z}^{(i)}-\frac{\gamma}{2pN^{p-1}}\sum_{i_{1},i_{2},...,i_{p}=1}^{N} \sigma_{x}^{(i_{1})}\sigma_{x}^{(i_{2})}... \sigma_{x}^{(i_{p})}\\ &=-hJ_{z}-\frac{\gamma}{pJ^{p-1}}J_{x}^{p}
    \end{split}
\end{align}
where $h$ is the strength of the external field, $\gamma$ is the strength of the $p$-body Ising-like interaction, and $J_\mu = \frac{1}{2}\sum_{i=1}^{N} \sigma_{\mu}^{(i)}$ with $\mu = x,y,z$, are collective spin operators. The interaction term has been normalized with this particular choice of $p$ and $N$ to make the equations of motion have a universal form for all $p$ in the mean-field limit and energy extensive~\cite{Munoz2020}. In this Hamiltonian, the total angular momentum is conserved, $[H,J^{2}]=0$, constraining the dynamics to the symmetric subspace, and the 
system is also invariant under the action of the parity operator, $\Pi=e^{i\pi J_{z}}$, for even $p$ spin models. For the remainder of this manuscript,  we consider the following single-parameter version of the $p-$spin Hamiltonian,
\begin{align}
    \label{eqn:one_param_p_spin_hami}
    H(s)=-(1-s) J_{z}-\frac{s}{pJ^{p-1}}J_{x}^{p},
\end{align}
which is equivalent to Eq. (\ref{eqn:p_spin_hami}) upon rescaling of the energy, and where $s$ is constrained to be in the range $0 \leq s \leq 1$, interpolating between pure paramagnetic ordering and a pure ferromagnetic ordering.  For $p=2$, the above Hamiltonian reduces to the Curie-Weiss model~\cite{Kochmanski2013}, which is a special instance of the Lipkin-Meshov-Glick (LMG) model~\cite{Lipkin1965,Santos2016,Castanos2006}. This family of models has been extensively studied in the context of quantum annealing \cite{Jorg2010,matsuura2017}, where a classification based on the properties of the ground state phase transition (GSQPT) was constructed~\cite{Filippone2011,Bapst2012}. Such classification splits the family of models in two classes, for $p=2$ the GSQPT is continuous and second order, for $p>2$ the GSQPT is first order and discontinuous. Furthermore, in the context of dynamical criticality, this family of models exhibit dynamical quantum phase transitions \cite{Munoz2020}.

The dynamics of the $p$-spin Hamiltonians in the mean-field limit can be obtained by neglecting the fluctuations $\langle A B \rangle \approx \langle A \rangle \langle B \rangle$ in the Heisenberg equations of motion resulting in the coupled differential equations of the form \cite{Milburn1997,Munoz2020}
\begin{align}
\label{eq:class_eqns_p_spin}
\begin{split}
\frac{dX}{dt}	&=(1-s)Y , \\
\frac{dY}{dt}	&=-(1-s)X+s X^{p-1}Z ,\\
\frac{dZ}{dt}	&=-sX^{p-1}Y .\\
\end{split}
\end{align}
with $\{X,Y,Z\}=\lim\limits_{J\rightarrow \infty}\frac{1}{J}\{\langle J_{x} \rangle,\langle J_{y} \rangle,\langle J_{z} \rangle\}$, describing the motion of a classical ``top''.  The mean-field limit coincides with the classical limit, $\hbar_{\rm eff}=N^{-1} \rightarrow 0$ or equivalently $N\rightarrow \infty$, in the case of $p$-spin models \cite{Bapst2012}. Due to the conservation of angular momentum, the resulting dynamics of the top is constrained to a unit sphere $X^{2}+Y^{2}+Z^{2}=1$. This implies that all the $p$-spin models are integrable in the classical limit as they correspond to autonomous systems with one degree of freedom. 

The first-order Trotterized unitary map $U_{\rm trot}=(U_{\delta}(\tau))^{n}$  of the Hamiltonian evolution in Eq. (\ref{eqn:one_param_p_spin_hami}) is generated by the time-evolution of the corresponding kicked model,
\begin{equation}
\label{eq:one_param_kick_p_spin_hami}
 H_{\delta}(t)=-(1-s) J_{z}-\frac{s\tau}{pJ^{p-1}} \sum_{n=-\infty}^{\infty}\delta \bigl(t-n\tau\bigr)J_{x}^{p}.\\
\end{equation}
We refer to this family of models as the kicked $p$-spin models \cite{Munoz2021}, whose Floquet operator is given by 
\begin{align}
\label{eq:kicked_spin}
    U_{\delta}(\tau)=e^{i(1-s)\tau J_{z}} e^{i\frac{s\tau}{pJ^{p-1}} J_{x}^{p}}\equiv F(\tau).
\end{align}
The equations of motion are then obtained using the map $\textbf{J}_{i+1}=F(\tau)^{\dagger} \textbf{J}_{i} F(\tau)$. Note that these models also conserve the angular momentum, $[H_{\delta},J^{2}]=0$ similar to the case of $H_{\text{tar}}$ implying that the dynamics of $H_{\delta}$ is constrained to a unit sphere defined by $X_{i}^{2}+Y_{i}^{2}+Z_{i}^{2}=1$ in the classical limit. The classical equations of motion are given by the following map,
\begin{align}
\label{eq:class_eqns_kicked_p_spin}
\begin{split}
    X_{m+1}&=X_{m}\cos{\alpha}-Y_{m}\sin{\alpha} ,\\
    Y_{m+1}&=\cos{[k(X_{m+1})^{p-1}}](X_{m}\sin{\alpha}+Y_{m}\cos{\alpha})\\
    &\qquad\qquad\quad\qquad -Z_{m}\sin{[k(X_{m+1})^{p-1}}],\\
    Z_{m+1}&=\sin{[k(X_{m+1})^{p-1}}](X_{m}\sin{\alpha}+Y_{m}\cos{\alpha})\\
    &\qquad\qquad\quad\qquad +Z_{m}\cos{[k(X_{m+1})^{p-1}}],
\end{split}    
\end{align}
where $\alpha =-(1-s)\tau$ and $k=-s\tau$. A comparison of targeted Hamiltonian flow, Eq. (\ref{eq:class_eqns_p_spin}), at periodic intervals with the area-preserving map generated by the Trotterized kicked Hamiltonian, Eq. (\ref{eq:one_param_kick_p_spin_hami}), indicates where errors may occur in the quantum simulation.

The kicked models can exhibit chaos since the energy of the top is no longer conserved. For small values of $\tau$, however, where the kicked-systems in Eq. (\ref{eq:kicked_spin}) very closely approximate the evolution of corresponding Hamiltonians in Eq. (\ref{eqn:one_param_p_spin_hami}), the dynamics is regular. As the value of $\tau$ is increased, all kicked $p$-spin models become chaotic provided $s$ is not close to zero or one; the models with larger value of $s$ develop chaos at smaller values of $\tau$. The kicked $p=2$ model (Haake's kicked-top~\cite{haake1987}) develops chaos due to the period-doubling cascade and reaches the regime of strong chaotic trajectories faster (as a function of the coefficient of non-linear term in the Hamiltonian) than other $p$-spin models \cite{Munoz2021}. Higher-order kicked $p$-spin models, on the other hand, develop chaos due to instability of higher-period orbits present on the $XY$ plane \cite{Munoz2021}. 

In order to analyze the emergence of quantum simulation errors in the different parameter regimes of the Trotterized evolution, we perform a systematic comparison between the basis of eigenstates of $U_{\text{tar}}$ and $U_{\text{trot}}$. In Figs. 1a and 1b, we plot the average dissimilarity between both sets of eigenstates for the $p$-spin model with $p=2$ and $p=4$, correspondingly. This quantity measures the difference between two sets of eigenvectors and is defined as
\begin{align}
\label{eq:dissimilarity_definition}
    \rm D(U_{\rm tar},U_{\delta})=\frac{1-\overline{\rm IPR}}{1-\overline{\rm IPR}_{\text{COE}}},
\end{align}
where $\overline{\rm IPR}$ is the average inverse participation ratio (IPR) of the eigenstates of $U_{\rm tar}$, denoted by $\{|\phi_{\rm tar}^{(j)}\rangle\}$, in the eigenstates of $U_{\delta}$, $\{|\phi_{\delta}^{(i)}\rangle\}$, which is given by
\begin{align}
\label{eq:IPR_definition}
\overline{\rm IPR}=\frac{1}{d}\sum_{i,j}|\langle \phi_{\rm tar}^{(i)}|\phi_{\delta}^{(j)}\rangle|^{4},
\end{align}
where $d$ is the dimension of the Hilbert space. Also, $\overline{\rm IPR}_{\text{COE}}=\frac{3}{N+3}$ is the average IPR in the circular orthogonal ensemble (COE) \cite{siberer2019}. We expect the average IPR between the eigenstates of $U_{\text{tar}}$ and $U_{\text{trot}}$ to be equal to $\overline{\rm IPR}_{\text{COE}}$ when the dynamics of $U_{\text{trot}}$ becomes fully chaotic, as COE is the appropriate ensemble due to the symmetries present in the kicked $p$-spin Hamiltonian \cite{Munoz2021}. The dissimilarity defined in Eq. (\ref{eq:dissimilarity_definition}) ranges between $0$, when the eigenstates are identical to the reference basis, and $\frac{d-1}{d}$ (up to normalization) for the case when the eigenstates are completely delocalized in the reference basis. 

The dissimilarity of the eigenvectors is shown in the heat map in Fig. \ref{fig:fig1}a and \ref{fig:fig1}b.  At larger values of $s$, the dissimilarity arises mainly due to the chaos present in the kicked $p$-spin models.  This is indicated by the region of parameter space with a positive classical Lyapunov exponent (see in Fig. 1c and 1d for $p=2$ and $p=4$, correspondingly. For more details on calculation of Lyapunov exponents, see Appendix \ref{app:Lyapunov} and \cite{Constantoudis1997,Munoz2021}). 
In contrast, the large dissimilarity present at smaller values of $s$ on the heat map cannot be attributed to chaos since the associated Lyapunov exponents are zero. We identify these parameter regimes as {\em structural instabilities} of the operator $U_{\text{trot}}$, since small changes in the parameter space $(s,\tau)$ induce substantial changes in the nature of the eigenstates of the operator. In the following section, we will derive the precise location of these structural instabilities, and relate them to the proliferation of errors in quantum simulation. 

\section{Trotter errors in observables due to structural instabilities}
\label{sec:structural_instabilities}
\begin{figure}[t]
\centering
\includegraphics[width=\linewidth]{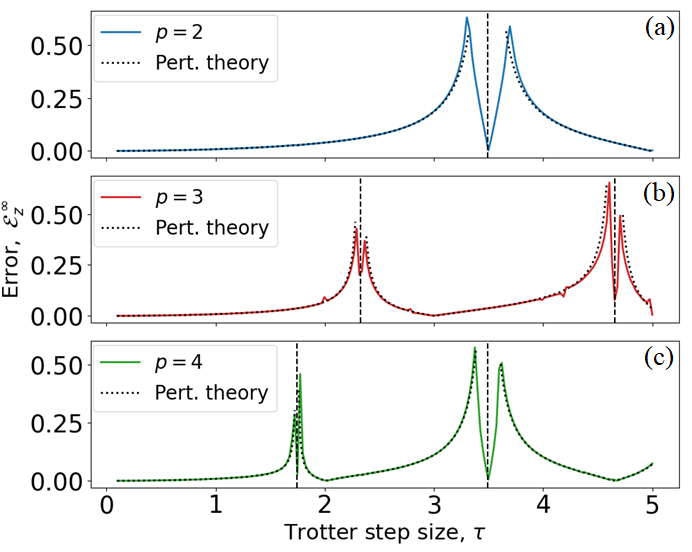}
\caption{Error in the long-time averaged magnetization along the $z$-axis, denoted by $\mathcal{E}_{z}^{\infty}$ and defined in Eq. (\ref{eq:error_definition}), for $p=2$, $3$ and $4$ in Figs (a), (b) and (c) respectively for $N=256$. The dotted lines represent time-averaged magnetization predicted from the nondegenerate first-order unitary perturbation theory, which agree well with numerically obtained values except in the immediate vicinity of the center of the structural instability region where the non-perturbative term $e^{i(1-s)\tau J_{z}}$ becomes degenerate. The vertical dashed lines are located at the center of structural instability regions, as predicted by the unitary perturbation theory.}
\label{fig:fig2}
\end{figure}
In this section we predict the location of regions of structural instability identified in Fig. \ref{fig:fig1} and analyze their effect on quantum simulation errors. In order to do this, we focus on the weakly interacting regime of the $p$-spin Hamiltonian, corresponding to small values of the coupling parameter, $s\ll 1$. At small values of $s$, the Floquet map implements precessions around the $z$-axis with small perturbation, $U_{\delta}=e^{i(1-s)\tau J_{z}}e^{i\frac{s\tau}{pJ^{p-1}} J_{x}^{p}}\equiv U^{(0)}U^{'}$. Hence the eigenvectors of the Floquet operator $U_{\delta}$ are expected to be close to the eigenstates of $J_{z}$ with a small correction first order in $s$. Employing unitary perturbation theory~\cite{Peres2006}, we find 
\begin{align}
\begin{split}
        U_{\delta}\biggl(|-J+m\rangle+s |\phi_{m}^{(1)}\rangle \biggr)&=e^{i(1-s)\tau(-J+m)}(1+is\phi^{(1)}_{m}) \\
        &\qquad\qquad \biggl(|-J+m\rangle+s |\phi_{m}^{(1)}\rangle \biggr).
\end{split}
\end{align}  
Here $|-J+m\rangle \equiv |J,m_{z}=-J+m\rangle$ is the zeroth order $m^{th}$ eigenstate of $J_z$, $\phi_{m}^{(1)}$ is the first-order eigenphase correction and $|\phi_{m}^{(1)}\rangle$ is the corresponding first-order  correction
to the eigenstate. This can be expressed as 
\begin{equation}
\label{eq:eigenvector_correction}
\langle -J+m'|\phi_{m}^{(1)}\rangle=(1-\delta_{m',m})\frac{i\tau}{pJ^{p-1}}\frac{\bigl(J_{x}^{p}\bigr)_{m',m}}{e^{i(1-s)\tau(m-m')}-1},
\end{equation} 
where $\bigl(J_{x}^{p}\bigr)_{m',m} \equiv \langle -J+m'|J_{x}^{p}|-J+m\rangle$ and $m,m'=\{0,1,...,2J\}$.  

Note that the ideal unitary map, $U^{(0)}$, has degenerate eigenvalues when $e^{i(1-s)\tau(m-m')}=1$ (equivalently at $\tau = r\frac{2\pi}{(1-s)(m-m')}$ for some positive nonzero integer, $r$). We denote the degenerate points by $\tau^{*}_{p,m-m'}$ corresponding to the kicked $p$-spin model with a particular solution for $m-m'$.
The neighborhood of these degenerate points can be divided into two subregions. First, in the zone surrounding the degenerate point, the first-order correction to the eigenphase is on the order of gap between the eigenphases, and the nondegenerate perturbation theory is not valid in this subregion. We refer to this subregion as the 
``immediate vicinity" of the degenerate point. In the subregion beyond this immediate vicinity (``outer vicinity"), the non-degenerate perturbation theory is valid, and the expression in Eq. (\ref{eq:eigenvector_correction}) predicts a large correction to the eigenstates whenever $(J_{x}^{p})_{m',m}\neq 0$. We refer to the whole region consisting of immediate and outer vicinity subregions surrounding the degenerate points as the structural instability regions whenever $(J_{x}^{p})_{m',m}\neq 0$. We will refer to $\tau^{*}_{p,m-m'}$ as the structural instability points, which are the central points of these regions, and denote width of the regions by $w$.

For concreteness, consider the area around $\tau=\tau^{*}_{2,2}=r\frac{\pi}{(1-s)}$, that is, corresponding to $p=2$ and $m-m'=2$. In this case, we have that $\bigl(J_{x}^{2}\bigr)_{m',m} \neq 0$  while $e^{i(1-s)(m-m')\tau}-1$ approaches $0$ in the neighborhood of $\tau=\tau^{*}_{2,2}$. Thus, by virtue of Eq. (\ref{eq:eigenvector_correction}), we obtain a substantial change in the eigenstates of the unitary evolution operator, indicating the existence of an instability region. At $\tau=\tau^{*}_{2,2}$, the Floquet operator is given by $U_{\delta}(\tau^{*}_{2,2})=e^{i\pi J_{z}}e^{i\frac{s\tau^{*}}{2J}J_{x}^{2}}$ whose eigenstates are $\bigl\{\frac{1}{\sqrt{2}}(|J,m_{x}\rangle\pm |J,-m_{x}\rangle)\bigr\}$. This can be understood from the fact that the two terms in $U_{\delta}(\tau^{*}_{2,2})$ commute and therefore have a common set of eigenvectors given by the parity respecting eigenstates of $J_{x}^{2}$. Hence, as the Trotter-step size is varied in the vicinity of $\tau^{*}_{2,2}=r\frac{\pi}{(1-s)}$, the eigenstates of $U_{\delta}$ change rapidly from those of $J_{z}$ eigenstates at the left edge of the structural instability region ($\tau \sim \tau^{*}_{2,2}-
\frac{w}{2}$) to those of parity respecting eigenstates of $J_{x}^{2}$ at the central point of the instability region before changing back to those of $J_{z}$ eigenstates as $\tau$ is increased further towards the other edge of the instability region ($\tau \sim \tau^{*}_{2,2}+\frac{w}{2}$).
The above argument holds for all even $p$-spin models, so all these models have structural instability regions in the vicinity of the curve $(1-s)\tau^{*}_{p,2}=r\pi$. 

In general, for a given $p$, $J_{x}^{p}$ has nonzero matrix elements in the $J_{z}$-basis only in alternating diagonal bands up to offset $p$.  This results in structural instabilities in the vicinity of the curves given by $(1-s)\tau^{*}_{p,m-m'}=\frac{r}{m-m'} 2\pi$  with $m-m'=\{p,p-2,p-4,...,2(1)\}$ for even (odd) values of $p$. We corroborate this in Fig. \ref{fig:fig1}a and \ref{fig:fig1}c, where we show the curves $(1-s)\tau^{*}=\pi$ (black-dashed line) and $(1-s)\tau^{*}=\frac{\pi}{2}$ (black-dotted line) overlap with the dissimilarity region present on the heat map. Note that the number of structural instability regions increases with $p$ in a given range of $\tau$ as the number of choices for $m-m'$ increase with $p$.  

The significant change in the eigenstates of $U_{\delta}$ that we find in the structural instability regions implies that $U_{\delta}$ becomes very different from $U_{\rm tar}$, which can lead to large errors in a Trotterized quantum simulation algorithm. For concreteness, we focus on the simulation of the long-time average of the collective spin observables $\langle J_{i} \rangle$, defined by 
\begin{align}
    \overline{\langle J_{i} \rangle}=\lim\limits_{n\rightarrow \infty}\frac{1}{n}\sum_{l=1}^{n} \langle J_{i}(l\tau) \rangle ,
\end{align}
where $J_{i}(l\tau)=(U^{\dagger})^{l}J_{i}(U)^{l}$, with $U$ being the map associated with the time-evolution operator for $t=\tau$ and $i=\{x,y,z\}$.
We analyze the error in $\overline{\langle J_{z} \rangle}$ 
given by
\begin{align}
\label{eq:error_definition}
    \mathcal{E}_{z}^{\infty}(\tau)=\frac{1}{J}\bigl| \overline{\langle J_{z}\rangle }_{\rm tar} - \overline{\langle J_{z}\rangle}_{\rm trot} \bigr|
\end{align}
where $\overline{\langle J_{z}\rangle }_{\rm tar}$ and  $\overline{\langle J_{z}\rangle }_{\rm trot}$ are the time-averaged magnetizations obtained under the target unitary and the Trotterized unitary correspondingly. 
As studied in~\cite{Poggi2020, siberer2019}, quantum simulation is expected to be robust to imperfections in the nonchaotic regime of $H_{\delta}$ for expectation values of macroscopic observables that are not sensitively dependent on the full state of the system compared to quantities such as the fidelity of preparing a target state. However, the Trotterization of $p$-spin models leads to a large region of error, even in the simulation of a macroscopic quantity such as $\overline{\langle J_{z}\rangle}$. In Fig. \ref{fig:fig2} we have plotted the error  $\mathcal{E}_{z}^{\infty}(\tau)$ for an initial spin coherent state $|\Theta_{0}=\frac{\pi}{2},\Phi_{0}=0\rangle$ at $s=0.1$ as a function of the Trotter-step size for $p=2,3$ and $4$ in parts (a),(b) and (c) respectively. As expected, the errors increase in structural instability regions in the vicinity of $\tau^{*}_{2,2}=\frac{\pi}{1-s}$ for $p=2$, $\tau^{*}_{3,3}=\frac{2\pi}{3(1-s)}$, $\tau^{*}_{3,1}=\frac{2\pi}{(1-s)}$ for $p=3$ and $\tau^{*}_{4,4}=\frac{\pi}{2(1-s)}$, $\tau^{*}_{4,2}=\frac{\pi}{(1-s)}$ for $p=4$ cases (all cases shown by the vertical dashed lines). 
Note that at every structural instability region shown in Fig. \ref{fig:fig2}, the errors first increase rapidly, then decrease in some intermediate region before increasing again resulting in seemingly two separate error peaks. The presence and location of the error dip between the error peaks is dependent on the initial condition, and will be analyzed in the next section. 

An important consequence of this analysis is that the structural instability regions associated with higher values of $p$ occur at smaller values of $\tau$ for a given value of $s$, since
$\tau_{p,m-m'}^{*}=r\frac{2\pi}{(1-s)(m-m')}$ where $m-m'=\{p,p-2,...,2(1)\}$. As the kicked $p$-spin models become chaotic at larger values of $s$ (for moderate values of $\tau$), these instabilities are the only source of errors for Trotterized simulation at smaller values of $s$, which corresponds to a regime where the external field in Eq. (\ref{eq:one_param_kick_p_spin_hami}) dominates over the multi-spin interaction.

The spectral gap given by the denominator term in Eq. (\ref{eq:eigenvector_correction}) determines the width of the error regions. From the Taylor expansion of the inverse spectral gap around the structural instability points $\tau=\tau^{*}$, we have, 
\begin{align}
    \frac{1}{e^{i(1-s)\tau(m-n)}-1}\approx \frac{e^{-i(1-s)\tau(m-n)}-1}{(m-n)^{2}(1-s)^{2}(\tau-\tau^{*})^{2}}.
\end{align}
The parabolic form of the denominator centered at $\tau^{*}$ with the width given by $\frac{1}{(m-n)^{2}(1-s)^{2}}$ implies that error regions become narrower for larger values of $m-m' \;\bigl(=\{p,p-2,...\}\bigr)$ and smaller values of $s$. This explains the fact that the error regions around $\tau=\tau_{p,m-m'}^{*}$ with $m-m'=2\; (\text{or}\; 1 \; \text{for odd p-spin models})$ are the widest and become narrower as $m-m'$ corresponds to higher values.

The error in the long-time averaged magnetization along the $z$-axis to first-order in $s$ for initial spin-coherent states  $|\Psi^{(0)}\rangle=|\Theta,\Phi\rangle $ can be expressed analytically.  We  find,
\begin{widetext}
\begin{align}
\label{eq:error_expression}
\begin{split}
\mathcal{E}_{z}^{\infty}(\tau)=\biggl|\sum_{q=\{p,p-2,...,2(1)\}}\frac{sq}{pJ^{p-1}}\biggl[\cos(q\Phi)\biggl(\frac{2}{q(1-s)} -\tau\cot\bigl(\frac{q(1-s)}{2}\tau\bigr)\biggr)+\tau\sin(q\Phi)\biggr] \sum_{m=0}^{2J-q} |\rho^{(0)}_{m+q,m}(\Theta)|\;(J_{x}^{p})_{m,m+q} \biggr|,
\end{split}
\end{align}
\end{widetext}
where $A_{r_{1},r_{2}}=\langle -J+r_{1}|A|-J+r_{2} \rangle$ for an operator $A$. The error expression for an arbitrary initial state is shown in the Appendix \ref{appendix:error_expression}. Equation (\ref{eq:error_expression}) predicts an error peak, captured by the cotangent term in the outer vicinity region, for each value of $q$ in the summation, which corresponds to having an error peak at every structural instability region. This result from perturbation theory also agrees very well with the numerically obtained curves as shown by the dotted lines in Fig. \ref{fig:fig2} for the spin-coherent states centered at $|\Theta=\frac{\pi}{2},\Phi=0\rangle$ except in the immediate vicinity of the degenerate point, where the errors have inverted-triangular shape (analytic prediction in this region is not shown in Fig. \ref{fig:fig2} because these predictions diverge here, and we expect this because the non-degenerate perturbation theory is not valid in this region). This behavior holds true for most of the other spin-coherent states at small values of $s$.

\section{Effective Hamiltonian and emergent symmetries}
\label{sec:effective_hamil}
\begin{figure*}[t!]
\centering
\includegraphics[width=\textwidth]{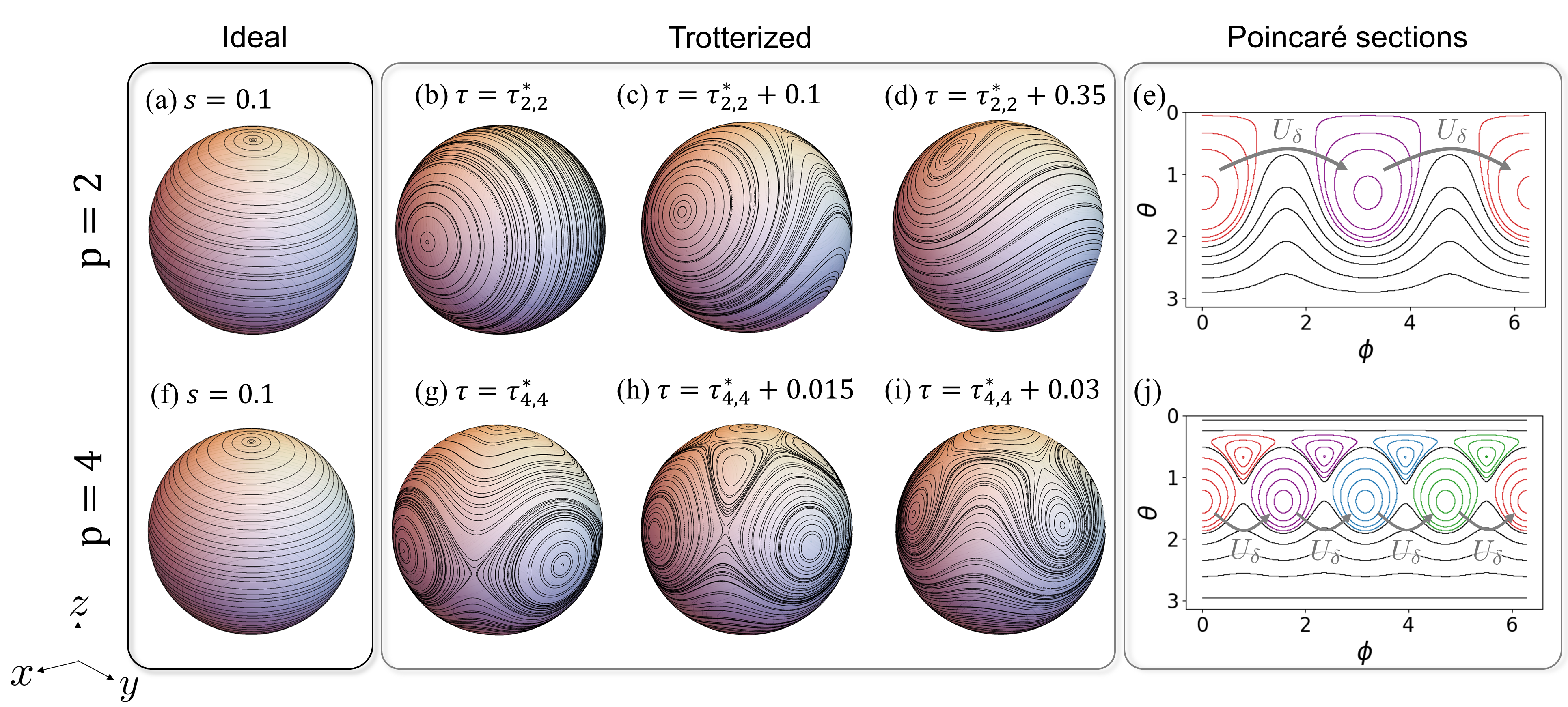}
\caption{\textbf{(a)}: Classical phase-space trajectories 
associated with the mean-field dynamics of target unitary map for $p=2$ (LMG model) at $s=0.1$.\textbf{(b)-(d)}: Classical phase space associated with the Trotterized unitary for $p=2$ at $s=0.1$ for various Trotter-step sizes located in the structural instability region, centered at $\tau=\tau^{*}_{2,2}=\frac{\pi}{0.9}$. The phase-space trajectories in (b) through (d) show that the targeted simulation of the paramagnetic phase dynamics through Trotterization results in simulation of the ferromagnetic phase of the LMG model when the Trotter-step sizes are chosen around $\tau^{*}_{2,2}$. This process is accommodated by a $1$-to-$2$ bifurcation at $Z=1$ \textbf{(e)}: Identical to the phase space shown in part (c) but plotted as a function of polar and azimuthal angle. Parity-broken trajectories are colored red and purple to illustrate that states initialized in the lobed region jump between red colored curve and the corresponding purple colored curve tracing out two trajectories at the same time. On the other hand, parity broken trajectories (black-color) trace out the actual LMG trajectories. \textbf{(f)}: Same as in part (a) except this is for $p=4$. \textbf{(g)-(i)}: The simulated Hamiltonian is now given by $H^{(4,4)}_{\rm eff}=-(1-s)\frac{\Delta \tau}{\Delta \tau+\tau^{*}}J_{z}-\frac{s}{8J^{3}}(J_{x}^{4}+J_{y}^{4})$, and the point $Z=1$ has two separate $1$-to-$4$ bifurcations in the instability region for $\Delta \tau >0$. \textbf{(j)}: Phase space shown in part (h) is plotted as a function of polar and azimuthal angle. As a result of $1$ to $4$ bifurcations, a state initialized in one of the lobes jumps between different lobes colored red, purple, blue and green and traces out four different trajectories (one in each lobe)}
\label{fig:fig3}
\end{figure*}

As seen in the previous section, Trotterization of the $p$-spin models leads to large errors in the vicinity of certain parameter regimes corresponding to so-called structural instabilities. This was understood in the classical limit to be a consequence of bifuractions occurring in the area-preserving map of the Trotterized model that radically shift the structure of phase space, and which manifests at the quantum level in a Floquet operator whose eigenstates are very different from those of the target $p$-spin Hamiltonian. This difference in the structure of eigenstates can be further elucidated through the construction of an effective Hamiltonian associated with the Trotterized unitary. For small values of $s$, we have $U_{\rm tar} \sim e^{i\kappa J_{z}}$, and the evolution is essentially precession of states around the $z$-axis. These precessions are well approximated by the Trotterized evolution $U_{\rm trot}=U_{\delta}(\tau)^{n}$ away from the structural stabilities. However, near these instabilities, the phase space of the Trotterized evolution
undergoes major structural changes leading to a evolution very different from precessions of the state around the $z$ axis.

For instance, consider $p=2$, the LMG Currie model, whose phase space in the classical limit associated with $U_{tar}$ is shown in Fig. \ref{fig:fig3}(a). $U_{\delta}(\tau)$ is shown in Fig. \ref{fig:fig3}(b-d) at $s=0.1$ in the neighborhood of the instability that is present at $\tau^{*}_{2,2}=\frac{\pi}{1-s}\approx 3.49$. We understand this structure of phase space by analyzing the form of $U_{\delta}^{2}$,
\begin{align}
\label{eq:eff_unitary_p2}
\begin{split}
    \bigl(U_{\delta}(\tau^{*}_{2,2}+\Delta\tau)\bigr)^{2}&=\bigl(e^{i\pi J_{z}}e^{i(1-s)\Delta \tau J_{z}}e^{i\frac{s}{2J}(\tau^{*}_{2,2}+\Delta \tau) J_{x}^{2}}\bigr)^{2}\\
&=e^{i2\pi J_{z}} \bigl(e^{i(1-s)\Delta \tau J_{z}}e^{i\frac{s}{2J}(\tau^{*}_{2,2}+\Delta \tau) J_{x}^{2}}\bigr)^{2}.
\end{split}
\end{align}
For $s$,$\;\frac{\Delta \tau}{\tau^{*}_{2,2}+\Delta \tau}\ll 1$, the unitary map can be written as $\bigl(U_{\delta}(\tau^{*}_{2,2}+\Delta\tau)\bigr)^{2}=\pm \; e^{-i2(\tau^{*}_{2,2}+\Delta \tau)H^{(2,2)}_{\text{eff}}}$ with 
\begin{equation}
\label{eq:eff_hami_p2}
    H^{(2,2)}_{\rm eff}(s)=-(1-s)\frac{\Delta\tau}{\tau^{*}_{2,2}+\Delta \tau}J_{z}-\frac{s}{2J}J_{x}^{2},
\end{equation}
meaning that the dynamics for $p=2$ every two time steps can be described by the effective Hamiltonian $H^{(2,2)}_{\rm eff}$. For $\Delta \tau>0$, the effective Hamiltonian is in fact the LMG Hamiltonian of the form shown in Eq. (\ref{eqn:one_param_p_spin_hami}) with an additional overall multiplicative factor and an effective $s$ parameter given by
\begin{align}
    \frac{1-s_{\rm{eff}}}{s_{\rm{eff}}} \equiv \frac{(1-s)\frac{\Delta\tau}{\tau^{*}_{2,2}+\Delta \tau}}{s} \; ,
\end{align}
leading to
\begin{align}
\label{eq:s_effective}
s_{\text{eff}}=\frac{1}{1+\frac{1-s}{s}\frac{\Delta \tau}{\tau^{*}_{2,2}+\Delta \tau}}.    
\end{align}
$s_{\text{eff}}$ is always greater than $0.5$ in the region of the structural instability. This implies that the Trotter approximation of the unitary map with the original Hamiltonian having a {\em small} $s$ value (i.e., being in the paramagnetic phase) leads to simulation of the dynamics of the same model but with a {\em large} value of $s$, (i.e., corresponding to the  ferromagnetic phase, up to every alternate step). This paradoxical effect can also be seen in the classical phase space shown in \ref{fig:fig3}(b-d) for $\Delta \tau>0$, where the trajectories change from precessions around the $z$-axis for $\tau>\frac{\pi}{1-s}(1+\frac{s}{1-2s})\;(s_{\rm eff}>0.5)$ to precessions around $x$-axis at $\tau=\tau^{*}_{2,2}=\frac{\pi}{1-s}\;(s_{\rm eff}=1)$ as $\tau$ is decreased (from right to left in Fig. \ref{fig:fig3}).Notice that the time interval required to trace out the ferromagnetic phase dynamics using the Trotterized unitary (effective Hamiltonian) is $\frac{1}{s_{\rm{eff}}}$ slower compared to the time required for ideal LMG Hamiltonian ($p=2$ in Eq. (\ref{eqn:one_param_p_spin_hami})) with $s=s_{\rm{eff}}$ to follow the same effective dynamics. This can be traced back to the difference in the overall multiplicative factor,  $H_{\rm{eff}}^{(2,2)}(s)=\frac{1}{s_{\rm{eff}}} H(s_{\rm{eff}})$, where $H(s)$ is the Hamiltonian in Eq. (\ref{eqn:one_param_p_spin_hami}).

In the mean-field picture, this process is accommodated by a period-doubling bifurcation at $\tau=\frac{\pi}{1-s}(1+\frac{s}{1-2s})$ (corresponding to $s_{\rm eff}=0.5$),  where the stable fixed point at $Z=1$ becomes unstable as $\tau$ is decreased and a period $2$ orbit is created. As a result, even though the Trotterized phase space looks identical to the phase space for the LMG Hamiltonian with $s_{\text{eff}}$ given in Eq. (\ref{eq:s_effective}), the individual trajectories on the associated Trotterized phase-space trace out the LMG Hamiltonian trajectories only when one considers every alternate step of the Trotterized evolution. For example at $s=0.8$, the ideal LMG Hamiltonian described by Eq. (\ref{eqn:one_param_p_spin_hami}) with $p=2$ traces out both parity-broken trajectories (rotations), which are bounded by the separatrix, and parity-conserving trajectories (librations), and has a phase space that looks identical to the one associated with the Trotterized dynamics at $\tau=\tau^{*}+0.1 \approx 3.59$ and $s=0.1$ as shown in Fig \ref{fig:fig3}c. However, the Trotterized dynamics  trace out the two parity-broken trajectories, simultaneously as shown in Fig. \ref{fig:fig3}e, where the phase-space trajectories are plotted as a function of angular coordinates $\theta$ and $\phi$. For a given initial condition the Trotterized dynamics trace out one lobe (red-colored trajectories) in all the odd steps of the evolution and the other lobe in all the even steps (purple-colored trajectories) of the evolution. In this way, the Trotterized trajectory jumps between two separate parity-broken trajectories of the LMG Hamiltonian tracing out the ideal LMG dynamics with $s=s_{\rm eff}$ only every alternate step. On the other hand, for the initial conditions associated with the parity conserving trajectory of the LMG Hamiltonian, every step of the Trotterized unitary traces out the ideal LMG trajectory with $s=s_{\rm eff}$ (black-colored trajectories). Similar phenomenon takes place in the instability region for $\Delta \tau<0$ except that the bifurcation now takes place in the $Z<0$ hemisphere. 
Performing a similar analysis of the structural instability region present at $\tau^{*}_{p,2}=\frac{\pi}{1-s}$ for the even $p$-spin models shows that the Trotterized evolution also results in simulation of the ferromagnetic phase of the corresponding $p$-spin Hamiltonian (or the Hamiltonian with a relative negative sign for $\Delta \tau<0$) even though the target evolution is associated with the paramagnetic-phase dynamics.

The Trotterized unitary dynamics in the vicinity of other instabilities can differ even more substantially from the target, ideal dynamics. For example, consider the instability at $\tau^{*}_{4,4}=\frac{\pi}{2(1-s)}$ for $p=4$. The effective Hamiltonian can be derived in a similar manner as described above, yielding
\begin{equation}
    H^{(4,4)}_{\rm eff}=-(1-s)\frac{\Delta \tau}{\Delta \tau+\tau^{*}}J_{z}-\frac{s}{8J^{3}}(J_{x}^{4}+J_{y}^{4}),
\end{equation}
whose phase-space trajectories undergo two different $1$-to-$4$ bifurcations, as can be seen in Fig. \ref{fig:fig3}(g-i). The change in phase space structure results in trajectories that are very different from those associated with the target dynamics shown in Fig. \ref{fig:fig3}f. Similar to the case of $p=2$, the parity-broken trajectories of $H_{\rm eff}^{(4,4)}$ located on the phase-space are traced out by the Trotterized dynamics only every fourth step. In the intermediate steps, the Trotterized dynamics leads to jumps between various parity-broken lobes present on the phase space as shown in Fig. \ref{fig:fig3}j, where the red, purple, blue and green colored trajectories represent every first, second, third and fourth step respectively.

More generally, the dynamics in the vicinity of instability at  $(1-s)\tau^{*}_{p,q}=r\frac{2\pi}{q}$ has a $1$-to-$q$ bifurcation present on the classical phase of the Trotterized unitary and can be understood by analyzing $\bigl(U_{\delta}(\tau+\Delta \tau)\bigr)^{q}$. We see
\begin{align}
\begin{split}
    (U_{\delta}(\tau+\Delta \tau))^{q}&=\bigl(e^{i\frac{2\pi}{q}J_{z}}e^{i(1-s)\Delta \tau J_{z}} e^{i\frac{s}{pJ^{p-1}}(\tau^{*}_{p,q}+\Delta \tau)J_{x}^{p}}\bigr)^{q}\\
    &=\pm W_{\frac{(2q-2)\pi}{q}}...W_{\frac{4\pi}{q}}W_{\frac{2\pi}{q}}W ,
\end{split}
\end{align}
where
\begin{align}
    W \equiv e^{i(1-s)\Delta \tau J_{z}}e^{i\frac{s}{pJ^{p-1}}(\tau^{*}_{p,q}+\Delta \tau)J_{x}^{p}},
\end{align}
and
\begin{align}
    \begin{split}
    W_{\theta}&\equiv e^{-i\theta J_{z}}W e^{i\theta J_{z}}\\
    &=e^{i(1-s)\Delta \tau J_{z}}e^{i\frac{s}{pJ^{p-1}}(\tau^{*}_{p,q}+\Delta \tau)(J_{x}\cos \theta+J_{y}\sin \theta)^{p}}.
\end{split}
\end{align}
For $s$, $\frac{\Delta \tau}{\tau^{*}+\Delta \tau} \approx \frac{\Delta \tau}{\tau^{*}} \ll 1$, the Trotterized unitary can be expressed as $\bigl(U_{\delta}(\tau+\Delta \tau)\bigr)^{q}=e^{-iq(\tau^{*}_{p,q}+\Delta \tau)H_{\rm eff}^{(p,q)}}$ with the effective Hamiltonian given by
\begin{widetext}
\begin{align}
\begin{split}
    H^{(p,q)}_{\rm eff}=-(1-s)\frac{\Delta \tau}{\tau^{*}+\Delta \tau}J_{z}
    -\frac{s}{pqJ^{p-1}}\sum_{m=1}^{q}\biggl(J_{x}\cos\biggl[\frac{2\pi(m-1)}{q}\biggr]+J_{y} \sin\biggl[\frac{2\pi(m-1)}{q}\biggr]\biggr)^{p}.
    \end{split}
    \label{eq:heff}
\end{align}
\end{widetext}
The Hamiltonian in Eq. (\ref{eq:heff}) simplifies further when the Hamiltonian has parity symmetry, which is the case for all even $p$-spin models, as the $(\frac{q}{2}+k)^{th}$ term in the summation becomes identical to the $k^{th}$ term, reducing the number of terms in the summation from $q$ to $\frac{q}{2}$. This effective Hamiltonian captures the dynamics of every $q^{th}$ step of the Trotterized unitary in the vicinity of $1$-to-$q$ bifurcation. The associated phase-space of the effective Hamiltonian is invariant around the $z$-axis under $\frac{2\pi}{q}$ rotation since it commutes with $e^{i\frac{2\pi}{q}J_{z}}$: $[H_{\rm eff}^{(p,q)},e^{i\frac{2\pi}{q}J_{z}}]=0$. This is an emergent symmetry that appears in the structural instability region. We also want to point out that even though bifurcations facilitate the structural changes in the instability regions, not all bifurcations lead to such sharp changes in phase space. Only the subset of bifurcation points, identified here as “significant” bifurcations, lead to extensive changes in the structure of the eigenstates and result in large Trotter errors. These significant bifurcations appear only in the structural instability regions.

In summary, the effective Hamiltonian formulation explains the presence of large error peaks in the simulated time-averaged magnetization. At certain Trotter step sizes that correspond to structural instabilities, the Trotterized unitary evolution operator simulates a very different Hamiltonian from the target Hamiltonian. From a mean-field perspective, the major structural changes that take place inside the regions of structural instability always correspond with significant bifurcations on the classical phase space, which lead to creation of periodic hyperbolic points (hyperbolic fixed points of higher period), which are absent on the phase space associated with the target Hamiltonian $H_{\rm tar}\sim \kappa J_{z}$.

\section{Information scrambling inside structural instability regions}
\label{sec:inffo_scrambling}
\begin{figure}[t]
\centering
\includegraphics[width=\linewidth]{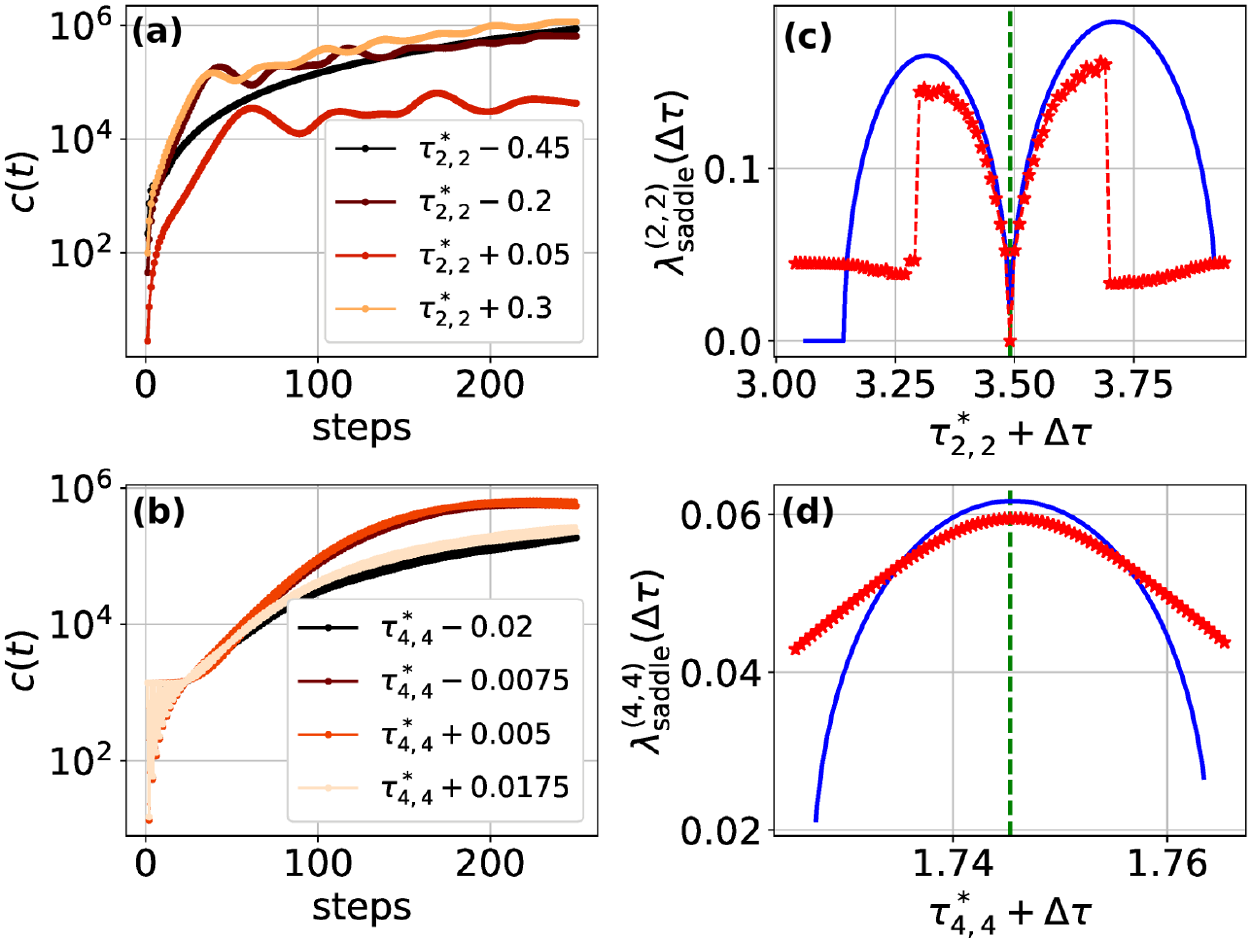}
\caption{\textbf{(a,b)} Examples of time evolution of the out-of-time-order correlation function (OTOC), $c(t)$, in Eq. (\ref{eq:our_square_commutator}) for some values of of $\Delta\tau$ inside the structural instability region for the $p=2$ \textbf{(a)} and $p=4$ \textbf{(b)}, and for the $\tau^*_{2,2}$ and $\tau^*_{4,4}$ structural instabilities, respectively. \textbf{(c,d)} Exponent of the saddle point controlling the rate of growth of the square commutator at short times. The solid line is the analytical prediction obtained from the classical flow associated with the effective Hamiltonian and the stars-dashed are the exponents obtained numerically by linear fit to the section of the data which grows exponentially. For this calculations, we used $N = 128$, $s=0.1$.}
\label{fig:fig4}
\end{figure}

As mentioned in the previous section, one of the signatures of the structural instability regions in the Trotterized unitary is the emergence of or unstable (hyperbolic) periodic fixed points. Thus, in the classical limit, the trajectories in the vicinity of this point are expected to show exponential divergence when the dynamics is observed at the appropriate stroboscopic times, \textit{i.e.}, every $q$ steps in the region around $\tau^*_{p,q}$. 

Recently, it has been shown that the presence of hyperbolic points can lead to information scrambling deep inside the quantum regime, a phenomenon that was dubbed saddle point scrambling~\cite{Xu2020,Kidd2021}. Note that this saddle-point scrambling is not indicative of chaotic behavior as mentioned in ~\cite{Xu2020,Kidd2021}, rather its origin lies in the exponential divergence of the trajectories that happens only in the localized region around the separatrix, which includes the hyperbolic fixed point. Here, we show that the effective Hamiltonian constructed in the previous section correctly identifies the presence of unstable fixed points and allows us to characterize the saddle point scrambling emerging in the simulator. Notice that such scrambling exists naturally in the dynamics of the ferromagnetic phase of the ideal $p$-spin models, as the existence of a critical (for $p=2$) or a bifurcation (for $p>2$) point is always accompanied by the emergence of an saddle point. This is easily understood as a pitchfork or a saddle-node bifurcation of the corresponding classical equations of motion. However, this type of scrambling is absent in the the paramagnetic phase, which is the target dynamics being simulated.  

The smoking gun of scrambling behavior is the exponential operator growth of certain correlation functions~\cite{Roberts2017}. In particular, we characterize the saddle-point scrambling in the quantum simulation using the short time growth of the ``infinite temperature'' square commutator 
\begin{equation}
\label{eq:square_commutator}
c(t)=\frac{1}{d}{\rm Tr}\left([V(t),W(0)] [V(t),W(0)]^\dagger \right) ,
\end{equation} 
where $d$ is the dimension of the Hilbert space, the operators $V(t)$ and $W(0)$ are chosen so that they commute at the initial time, and $V(t)$ is the Heisenberg evolution of $V(0)$. In this work, we choose $V(0) = W = J_{z}$, and study the short time growth of the square commutator
\begin{align}
\label{eq:our_square_commutator}
    c(t)=\frac{1}{N + 1}\text{Tr}\bigl(\bigl|[J_{z}(t),J_{z}(0)]\bigr|^{2}\bigr) .
\end{align}
In the presence of an instability, be it a saddle point or a hyperbolic periodic point, the above quantity is expected to grow like $c(t) \sim e^{\lambda_{\text{saddle}} t}$, where $\lambda_{\text{saddle}}$ is the associated growth rate at the saddle point. The exponential growth of the OTOC is seen in Fig. \ref{fig:fig4}a and \ref{fig:fig4}c for the $p=2$ system with Trotter step size in the vicinity of $\tau^{*}_{2,2}$ and $p=4$ system with Trotter step size around $\tau^{*}_{4,4}$, respectively. The exponents associated with the saddle point derived from these numerics are plotted in Fig. \ref{fig:fig4}b and \ref{fig:fig4}d as a function of the Trotter step size with the red dotted line. 

The analytic expressions for $\lambda_{\text{saddle}}^{(p,q)}$ associated with a given $p$ and $1$-to-$q$ bifurcation can be obtained by solving for the eigenvalues of the Jacobian matrix associated with the linearized classical flow around the appropriate unstable fixed point of the effective Hamiltonian, which are labelled by blue solid lines in Figs. \ref{fig:fig4}b and 4d.  In the system with $p=2$ around $\tau^*_{2,2}$, \textit{i.e} the $1$-to-$2$ bifurcation, one finds
\begin{equation}
\lambda^{(2,2)}_{\rm saddle} = 
    (\tau^*_{2,2} + \Delta\tau)\sqrt{\frac{s(1-s)|\Delta\tau|}{\tau^*_{2,2} + \Delta\tau} - \left(\frac{(1-s)\Delta\tau}{\tau^*_{2,2} + \Delta\tau}\right)^2}
\end{equation}
and the expression for $\lambda_{\rm saddle}$ in the system with $p=4$ around $\tau^*_{4,4}$, \textit{i.e} the $1$-to-$4$ bifurcation, is given by 
\begin{equation}
\lambda_{\text{saddle}}^{(4,4)} = \mathcal{M}_{+}^{(4,4)},
\end{equation}
where $\mathcal{M}_{+}^{(4,4)}$ is the largest eigenvalue of 
\begin{equation}
\mathbf{M}^{(4,4)} = \begin{pmatrix}
    0 & 2(1-s)\frac{\Delta\tau}{\tau^*_{4,4} + \Delta\tau} & \frac{s}{2}Y_{\rm sd}^3 \\
    -2(1-s)\frac{\Delta\tau}{\tau^*_{4,4} + \Delta\tau} & 0 & -\frac{s}{2}X_{\rm sd}^3 \\
    s Y_{\rm sd}^3 & -s X_{\rm sd}^3 & 0
\end{pmatrix},
\end{equation}
where $X_{\rm sd} = \frac{\sqrt{1 - Z^2_{\rm sd}}}{\sqrt{2}}$ and $X_{\rm sd}^2 = Y_{\rm sd}^2$ are the Cartesian coordinates of the unstable point of the classical flow associated with the effective Hamiltonian at this instability for the $p=4$ system. We give the explicit expression of $Z_{\rm sd}$ along with details on the derivation of the exponents in appendix \ref{app:exponents}.
In in Fig. \ref{fig:fig4}(b) and Fig. \ref{fig:fig4}(d) we compare the exponents extracted from the numerical calculation of the OTOC (red stars), for a system size of $N = 128$, with the analytical expressions obtained with the mean-filed limit of the effective Hamiltonian (solid blue). Notice the good agreement despite the small system size used in the simulation. Thus, as first observed in~\cite{Xu2020} the short time growth of the OTOC has the form $e^{\lambda_{\rm saddle} t}$. 


\section{Conclusions and outlook}
In this work we have identified a new mechanism leading to the proliferation of errors in a quantum simulator when the algorithm employs the Trotter-Suzuki decomposition. 
In the mean-field limit, these regions of structural instability are characterized by multiple bifurcations leading to rapid global changes in the structure of the phase space as the Trotter-step size is varied slightly. 

The effects of these bifurcations can be seen in smaller systems, with $N$ far from the thermodynamic limit. A method to identify the structural instability regions in these smaller systems is to seek the regions where the eigenstates of the Trotterized unitary differ significantly from the eigenstates of the target unitary. From the correspondence principle, the quasiprobability distribution is expected to overlap with the trajectories of the phase space in the classical limit, so rapid changes in the structure of the phase space are reflected in the modifications of the eigenstates associated with the Trotterized unitary. The sudden changes in the structure of the eigenstates of the Trotterized unitary result in Floquet dynamics that is very different from the targeted evolution of a given state, resulting in large errors in various observables. In this work, we provide analytic expressions for these high-error regions computed using unitary perturbation theory, for the case of $p$-spin models. We showed that inside the structural instability regions the effective Hamiltonian, which is the generator of the Trotterized evolution, is very different from the target Hamiltonian, providing further justification for the presence of large errors in these regions. The effective Hamiltonian reveals the emergence of new unstable fixed points in the structural instability regions, indicating the presence of saddle point scrambling in the simulator, as manifested by the exponential growth of the OTOCs in these regions.

An important conclusion of the perturbation theory analysis of Sec. \ref{sec:structural_instabilities} is that  structural instability regions will appear at smaller values of the Trotter step size $\tau$ as the value of $p$ increases. This indicates that, in general, gate-based quantum simulations of $p$-body interactions (beyond the usual two-body case of $p=2$) is likely to lead to more parameter regimes where Trotter errors proliferate. While this is certainly true in the case of all-to-all interactions  analyzed here, the study of its manifestation in systems with different interactions (i.e., with long, but finite, interaction range) will require further study. We conjecture that the structural instabilities studied here in the mean-field case are more universal, and carry over to Floquet states with more general many-body interactions. Note that models with all-to-all interaction graphs are often seen to correctly capture the physics of systems with more complex but still long-range interactions. For example, for the case of $p=2$, the mean-field phenomenology associated with the LMG model informs the DQPT behavior of finite-range interacting systems described by $H=B J_{z}+\sum_{i,j}\frac{J_{0}}{|i-j|^{\alpha}}  \sigma_x^{(i)} \sigma_x^{(j)}$ for $\alpha \lesssim 2$ ($\alpha=0$ corresponds to the LMG model) \cite{Zhang2017, Zunkovic2018}. These results strongly suggest that mean-field models can give insight into many-body behavior (particularly, the nonequilibrium dynamics of physical observables), and the notion of structural instabilities will very likely be present in finite-range models with nonzero values of $\alpha$ for the $p=2$ case. Further study of the structural instability regions in the finite-range interaction case, and the extension of these models for $p\geq 3$ is left for future work.

Beyond their implication in the proliferation of errors in quantum simulation, the regions of structural instabilities are a manifestation of fundamental effects in the nonequilibrium dynamics of the Floquet system. Particularly, we find that the dynamics of the driven system in Eq. (\ref{eq:kicked_spin}), whose Hamiltonian is periodic with period $T$, $H(t + T) = H(t)$,  shows signatures of Floquet time crystal behavior in a region of structural instability. A Floquet time crystal is an out-of-equilibrium phase of matter that breaks discrete time-translation symmetry \cite{Wilczek2015,Nayak2016,Russomanno2017,Yao2017}. These time-crystal phases in the kicked $p$-spin models can be studied with the help of area-preserving maps associated with the mean-field limit of the Floquet system. In particular, the existence of periodic elliptic points, which can be observed clearly in the phase space portraits of Fig. \ref{fig:fig3}, will lead to the dynamics where the initial state, prepared close enough to elliptic points, will periodically return to the initial configuration after $q$ time steps, with $q\geq 2$. Physically, this means that the response of the system will have a periodicity of $qT$, instead of $T$, thus breaking the discrete time-translation symmetry of the Hamiltonian. The connection between structural instabilities and subharmonic response was also seen in \cite{choudhury2021}, where the author studied the emergence of robust subharmonic response in a spin chain with short-range (nearest-neighbor) two-body interactions in a regime which roughly coincides with the choice of $\tau=\tau^*_{2,2}$ in the parametrization used in our work. This result, together with the general connection of the structural instability regions with discrete time crystals, indicates that this source of Trotter errors is not an isolated phenomenon happening only in long-range interacting spin models. A comprehensive study of all the Floquet time crystal phases present in the kicked $p$-spin system is part of an ongoing work and will be presented in an incoming publication~\cite{Munoz2021_crystals}.


\acknowledgments{We would like to thank Poul Jessen and Kevin Kuper for insightful discussions, and Changhao Yi for his insights in Trotter formulas and quantum simulation. This work was supported by the U.S. National Science Foundation under grant numbers PHY-1820679 and PHY-2011582. This material is based upon work supported by the U.S. Department of Energy, Office of Science, National Quantum Information Science Research Centers, Quantum Systems Accelerator.}

\pagebreak
\begin{widetext}
\appendix



\section{Calculation of Lyapunov exponents}
\label{app:Lyapunov}
Throughout the main text, we use the term Lyapunov exponent to describe the rate of exponential growth of a classical variable $X$. That is, up to some threshold time $t_{\rm th}$, this classical variable evolves according to $X(t) = X(0)e^{\Lambda t}$, with $\Lambda$ being the Lyapunov exponent. 

We have presented results in this work for two different scenarios where some components of the mean-spin in the thermodynamic limit evolve following the dynamics described in the previous paragraph. First, we have analyzed the chaotic instability of the delta-kicked $p$-spin as illustrated in Fig.~\ref{fig:fig1}c,d. Then, we analyze the short time evolution of the square commutator in Sec.~\ref{sec:inffo_scrambling}, whose physical origin can be traced to the exponential growth of the unstable separatrix branches of saddle points of the phase-space flow. Notice that we provide a more in-depth analysis of the instability around the saddle point in Appendix \ref{app:exponents}.


Although, both scenarios represent exponential instabilities, they correspond to physically different situations. The chaotic instability is global, so any pair of points inside the chaotic region of the phase space will display exponential divergence of their distance at a rate given by the exponent. The ubiquity of this instability all over the chaotic region leads to the folding of trajectories at long times, and the decay of correlations, the later known as mixing. In fact, at long times and length scales, the motion inside the chaotic region resembles a diffusion process. This implies that the chaotic dynamics has a positive value of metric entropy, or Kolmogorov-Sinai entropy, which can be computed via Pesin's theorem~\cite{Pesin1977}. For the instability of the saddle point, all of the above mentioned physical processes are absent. Furthermore, the exponential divergence of trajectories only occurs in a highly localized region of phase space including the immediate vicinity of the saddle point and the separatrix. For instance, see Ref.~\cite{Xu2020} for an in depth discussion on this matter.

The motivation behind Fig.~\ref{fig:fig1} is two fold. On the one hand, we want to introduce the disimilarity, and recognize that this quantity identifies both types of instabilities. On the other hand, we also want to present a direct comparison with the Lyapunov exponent of the chaotic instability, allowing us to identify the region of parameter space $(\tau,s)$ in the similarity heat map whose origin is chaos.

In the following, we explain the details of calculating the Lyapunov exponent associated with the chaotic instability, as shown in Fig.~\ref{fig:fig1}. In the thermodynamic limit, the mean-spin evolves stroboscopically according to Eq.~\ref{eq:class_eqns_kicked_p_spin}. For any point in phase space, the local dynamics of small increments in its vicinity is governed by the Jacobi matrix
\begin{equation}
\mathbf{M}(\mathbf{X}_m) = \frac{\partial \mathbf{X}_{m+1}}{\partial\mathbf{X}_m}.
\end{equation}
For a point inside the chaotic region, the exponential instability implies that the neighborhood of the point is getting exponentially strecthed in some of the principal directions of the Jacobi matrix and exponentially shrinked in the other principal directions. This takes place as one evolves the Jacobi matrix along the chaotic trajectory corresponding to the selected point. As such, the largest Lyapunov exponent can be computed as the largest eigenvalue of the Jacobi matrix evaluated along the trajectory, which follows from the celebrated Ergodic theorem of Oseledets~\cite{Ose1968,Eckmann1985}.

For a map as the one in Eq.~(\ref{eq:class_eqns_kicked_p_spin}), Oseledet's ergodic theorem allows us to compute the largest Lyapunov exponent given by 
\begin{equation}
 \label{eqn:oseledec}
 \Lambda_{+}(\tau, s; p) = \lim_{n\to\infty}[\lambda_{+}(\tau, s; p)]^{1/2n},
\end{equation}
where $n$ is the number of time steps, $\lambda_{+}$ is the largest eigenvalue of the matrix $\prod_{m=1}^N\mathbf{M}^T(\bm{X}_m)\mathbf{M}(\bm{X}_m)$ and $\mathbf{M}(\bm{X}_m)$ is the tangent map introduced before.

Naturally, the exponential growth of one of the eigenvalues leads to issues with the numerical computation of Eq.~(\ref{eqn:oseledec}). This can be avoided by looking at Eq.~(\ref{eqn:oseledec}) in a different basis other than the eigenbasis of the Jacobi matrix. This can be achieved via a QR decomposition, see for instance Ref.~\cite{Geist1990}, which permits the numerical approximation of the asymptotic time limit in Eq.~(\ref{eqn:oseledec}).

A single point in the heat maps of Fig.(\ref{fig:fig1})c,d is obtained by approximating Eq.~(\ref{eqn:oseledec}) via the QR method up to $n=10^6$ time steps. This value is then averaged over $50$ different initial points inside the chaotic region. This procedure is then repeated for a grid of points in the $(\tau, s)$ plane.

\section{Long-time average of $\langle J_{z} \rangle$}
\label{appendix:error_expression}
The long-time average of an operator $A$, assuming the time-evolution operator corresponding to one time-step has nondegenerate eigenphases, is given by 
\begin{align}
\overline{\langle A\rangle}_{\infty} &=\sum_{r=1}^{d} \langle \phi_{r}|\rho^{(0)}|\phi_{r}\rangle \langle \phi_{r}| A|\phi_{r}\rangle 
\end{align}
where $\rho^{(0)}$ is the initial state and $|\phi_{r}\rangle$ is the $r^{th}$ eigenstate of the system. The error in this observable due to Trotterization is given by
\begin{align}
    \mathcal{E}_{A}^{\infty}(\tau)= \frac{1}{J}|\overline{\langle A\rangle }_{\infty,\text{id}} - \overline{\langle A\rangle}_{\infty,\tau}|
\end{align}
where $\overline{\langle A\rangle }_{\infty,\text{id}}$ is the long-time average of $A$ under the ideal Hamiltonian evolution, and $\overline{\langle A\rangle}_{\infty, \tau}$ is the long-time average under Trotterized evolution.
Assuming that the eigenstates of the system change under a perturbation $|\phi_{r}\rangle \rightarrow |\phi_{r}^{(0)}\rangle + \lambda |\phi_{r}^{(1)}\rangle$, the expression for the long-time average to the first order is given by
\begin{align}
\label{eq:long_time_avg}
\begin{split}
&\overline{\langle A\rangle}_{\infty,\lambda}=\sum_{m=0}^{2J} \rho^{(0)}_{m,m} A_{m,m}+2\lambda\sum_{m,n\neq m}^{2J}\text{Re}\biggl(  \bigl(A_{m,m}\;\rho^{(0)}_{n,m}+\rho^{(0)}_{m,m}\; A_{n,m}\bigr)\langle \phi_{m}^{(1)}|\phi_{n}^{(0)}\rangle \biggr)
\end{split}
\end{align}
The above expression can be evaluated for ideal Hamiltonian evolution to obtain $\overline{\langle A\rangle}_{\infty,id}$ up to first order in $s$ using Hamiltonian perturbation theory with $H_{0}=-(1-s)J_{z}$ being the unperturbed Hamiltonian and $H_{1}=-\frac{s}{pJ^{p-1}}J_{x}^{p}$ being the perturbed Hamiltonian
\begin{align}
\begin{split}
 &\overline{\langle A\rangle }_{\infty,\text{id}}=\sum_{m=0}^{2J} \rho^{(0)}_{m,m} A_{m,m} +\frac{2s}{pJ^{p-1}(1-s)}\sum_{m,n\neq m}^{2J}\text{Re}\biggl(  \bigl(A_{m,m}\;\rho^{(0)}_{n,m}+\rho^{(0)}_{m,m}\; A_{n,m}\bigr)\frac{(J_{x}^{p})_{m,n}}{m-n} \biggr)
\end{split}
\end{align}
Likewise, Eq. (\ref{eq:long_time_avg})  can be evaluated to obtain $\overline{\langle A\rangle}_{\infty,\tau}$ up to first order in $s$ using unitary perturbation theory with unperturbed unitary, $U_{0}=e^{i(1-s)\tau J_{y}}$, and the perturbed unitary $U_{p}=e^{is\tau \frac{J_{z}^{p}}{pJ^{p-1}}}$
\begin{align}
\begin{split}
&\overline{\langle A\rangle}_{\infty,\tau}=\sum_{m=0}^{2J} \rho^{(0)}_{m,m} A_{m,m} + \frac{2s\tau}{pJ^{p-1}}\sum_{m,n\neq m}^{2J}\text{Re}\biggl(  \bigl(\rho^{(0)}_{n,m}A_{m,m}+A_{n,m}\rho^{(0)}_{m,m}\bigr) \frac{-i(J_{x}^{p})_{m,n}}{e^{-i(1-s)\tau(m-n)}-1}\biggr)
\end{split}
\end{align}
The error is then given by 
\begin{align}
\begin{split}
\mathcal{E}_{A}^{\infty}(\tau)=\frac{2s}{pJ^{p-1}}\biggl| \sum_{m,n\neq m}^{2J}& \text{Re}\biggl(  \bigl(A_{m,m}\;\rho^{(0)}_{n,m}+\rho^{(0)}_{m,m}\; A_{n,m}\bigr)(J_{x}^{p})_{m,n} \biggl(\frac{1}{(1-s)(m-n)} + \frac{i\tau}{e^{-i(1-s)\tau(m-n)}-1}\biggr)\biggr)\biggr|
\end{split}
\end{align}
The above expression can be further simplified by expanding the summation in $n$ and noticing the matrix elements of $(J_{x}^{p})_{m,n}$ are nonzero for $n={m\pm p, m\pm p-2,...,m \pm 0(1) }$. Focusing on two particular terms with $n=m\pm p-q$ we obtain
\begin{align}
\begin{split}
&\mathcal{E}_{A}^{\infty}(\tau)\bigl|_{n=m\pm q}=\frac{2s}{pJ^{p-1}}\sum_{m=0}^{2J-(p-q)}\text{Re}\biggl[  \biggl(\rho^{(0)}_{m+p-q,m}A_{m,m}+A_{m+p-q,m}
\qquad  \rho^{(0)}_{m,m}\biggr)(J_{x}^{p})_{m,m+p-q}\biggl(\frac{i\tau}{e^{i(p-q)(1-s)\tau}-1}-\frac{1}{(p-q)(1-s)}\biggr)\biggr] \\& \qquad +\frac{2s}{pJ^{p-1}}\sum_{m=p-q}^{2J}\text{Re}\biggl[\biggl(\rho^{(0)}_{m-(p-q),m}A_{m,m}+A_{m-(p-q),m}\rho^{(0)}_{m,m}\biggr)(J_{x}^{p})_{m,m-(p-q)}\biggl(\frac{1}{(p-q)(1-s)} + \frac{i\tau}{e^{-i(p-q)(1-s)\tau}-1}\biggr)\biggr]
\end{split}
\end{align}
Manipulating the second term in above expression by first shifting the index of the second term in the above equation, $m \rightarrow m-(p-q)$, and then setting $\text{Re}[z]=\text{Re}[z^{*}]$ in the second term results in the following expression for the error
\begin{align}
\begin{split}
\mathcal{E}_{A}^{\infty}(\tau)&=\frac{2s}{pJ^{p-1}}\sum_{q=\{0,2,...,p-1(p)\}}\sum_{m=0}^{2J-(p-q)}\text{Re}\biggl[  \biggl(\rho^{(0)}_{m+p-q,m}\bigl(A_{m+p-q,m+p-q}-A_{m,m}\bigr)\\&+A_{m+p-q,m}\bigl(\rho^{(0)}_{m+p-q,m+p-q}-\rho^{(0)}_{m,m}\bigr)\biggr) (J_{x}^{p})_{m,m+p-q}\biggl(\frac{1}{(p-q)(1-s)} -\frac{i\tau}{e^{i(p-q)(1-s)\tau}-1}\biggr)\biggr]
\end{split}
\end{align}
Focusing on the error in $J_{z}$, the above expression further simplifies to the following
\begin{align}
\begin{split}
\mathcal{E}_{z}^{\infty}(\tau)&=\frac{2s}{pJ^{p-1}}\sum_{q=\{0,2,...,p-1(p)\}}\sum_{m=0}^{2J-(p-q)}\text{Re}\biggl[\bigl(p-q \bigr) \rho^{(0)}_{m+p-q,m}(J_{x}^{p})_{m,m+p-q}
\biggl(\frac{1}{(p-q)(1-s)} -\frac{i\tau}{e^{i(p-q)(1-s)\tau}-1}\biggr)\biggr]
\end{split}
\end{align}
Relabelling the index $q \rightarrow p-q$ results in the final expression,
\begin{align}
\begin{split}
\mathcal{E}_{z}^{\infty}(\tau)&=\frac{2s}{pJ^{p-1}}\sum_{q=\{p,p-2,...,0(1)\}}\sum_{m=0}^{2J-q}\text{Re}\biggl[q \rho^{(0)}_{m+q,m}(J_{x}^{p})_{m,m+q}
\biggl(\frac{1}{q(1-s)} -\frac{i\tau}{e^{iq(1-s)\tau}-1}\biggr)\biggr]
\end{split}
\end{align}

\section{Details on the derivation of the growth rate of the square commutator}
\label{app:exponents}
As we mentioned in Sec. \ref{sec:effective_hamil} the exponent governing the growth rate of the square commutator can be obtained by examining the saddle points of the classical flow associated with the effective Hamiltonian constructed for a given power of the Floquet operator. Recall that, given that we are investigating the $q$th power of the Floquet operator, the effective Hamiltonian is given by 
\begin{equation}
    U_\delta(\tau)^q = e^{-iq(\tau^*_{p,q} + \Delta\tau)H_{\rm eff}}.
\end{equation}
Once the form of $H_{\rm eff}$ is known, the procedure is the following. First, we construct the equations of motion for the classical flow and identify its stationary points, particularly the unstable ones. Then, we evaluate the tangent map on this unstable points and compute its eigenvalues. The largest eigenvalue is then the exponent we are looking for.

\subsection{Structural instabilities around $\tau=\tau^*_{p,2}$}

\subsubsection{Structural instability around $\tau^*_{2,2}$}
We consider first the error peak arising due to the structural instability around $\tau^*_{2,2}$. For this system the effective Hamiltonian is given by 
\begin{equation}
    H^{(2,2)}_{\rm eff} = \frac{(1-s)^2}{\pi}\delta\tau J_z + \frac{s}{2J}J_x^2 = (1-s)\left(\frac{\Delta\tau}{\tau^*_{2,2} + \Delta\tau}\right)J_z + \frac{s}{2J}J_x^2.
\end{equation}
The equations of motion of the corresponding classical flow are given by
\begin{subequations}
\begin{align}
    \frac{dX}{dt} &= -\frac{(1-s)^2}{\pi}\Delta\tau Y, \\
    \frac{dY}{dt} &= \frac{(1-s)^2}{\pi}\Delta \tau X - s XZ, \\
    \frac{dZ}{dt} &= s X Y,
\end{align}
\end{subequations}
this classical flow has two fixed points at the poles, $X = Y = 0$ and $Z = \pm 1$. Other fixed points satisfy $Y = 0$ and 
\begin{equation}
Z = \frac{(1-s)^2 \Delta\tau}{\pi s} = \frac{1-s}{s}\frac{\Delta \tau}{\tau^*_{2,2} + \Delta\tau},\enspace\text{and}\enspace X = \sqrt{1 - Z^2}.    
\end{equation}
The range of values of $\Delta \tau$ for which these new fixed points are real gives us the extent of the region of structural instability, from the above expression for $X$ is easy to see that 
\begin{equation}
\Delta\tau \le \frac{s \tau^*_{2,2}}{1 - 2s},
\end{equation}
determines the width of the structural instability region.

Furthermore, it is not hard to see that the unstable point emerges as a consequence of the change in instability of one of the fixed points on the poles, depending on the sign of $\Delta\tau$. Hence, to compute the exponent we evaluate the Jacobi matrix on the poles and diagonalize it, finding that the two nonzero eigenvalues are given by
\begin{equation}
\mathcal{M}_\pm = \pm (1-s)\overline{\tau}\sqrt{{\rm sign}(\Delta\tau) \frac{s}{(1-s)\overline{\tau}} - 1},   
\end{equation}
where $\overline{\tau} = \frac{\Delta\tau}{\tau^*_{2,2} + \Delta\tau}$. From the largest eigenvalue we obtain the expression for the value of the exponent $\lambda_{\rm saddle}$  
\begin{equation}
\lambda^{(2,2)}_{\rm saddle}(\Delta\tau) = \Delta\tau (1-s)\sqrt{{\rm sign}(\Delta\tau) \frac{s}{(1-s)\overline{\tau}} - 1},
\end{equation}
where we have included the appropriate prefactor accounting for the definition of the effective Hamiltonian.

\subsubsection{Structural instability around $\tau^*_{4,2}$}
In Sec. \ref{sec:effective_hamil} we mentioned that all the even kicked $p$-spin models have a structural instability centered at the same value as the $p=2$ model, and the central value corresponds to the values of $\tau$ at which the period doubling bifurcation takes place. We now consider this structural instability region for the system with $p=4$. The effective Hamiltonian is given by 
\begin{equation}
 H^{(4,2)}_{\rm eff} = (1-s)\left(\frac{\Delta\tau}{\tau^*_{4,2} + \Delta\tau}\right)J_z + \frac{s}{4J}J_x^4.
\end{equation}
With the equations of motion for the associated classical flow given by 
\begin{subequations}
\begin{align}
    \frac{dX}{dt} &= -(1-s)\overline{\tau} Y, \\
    \frac{dY}{dt} &= (1-s)\overline{\tau}X - s X^3 Z, \\
    \frac{dZ}{dt} &= s X^3 Y,
\end{align}
\end{subequations}
where $\overline{\tau} = \frac{\Delta\tau}{\tau^*_{4,2} + \Delta\tau}$. This flow has two fixed points on the poles, $X = Y = 0$ and $Z = \pm 1$, which are always stable. New fixed points can be found as the solution to $\frac{d\bf{X}}{dt} = 0$. Of particular interest to us are the ones whose coordinates satisfy $Y = 0$ and the $x,z$-coordinates are related via 
\begin{subequations}
\begin{align}
(1-s)\overline{\tau} - s X^2Z &= 0, \\    
X^2 + Z^2 &= 1,
\end{align}
\end{subequations}
which leads to the cubic equation 
\begin{equation}
Z^3 - Z + \left(\frac{1-s}{s}\right)\overline{\tau} = 0.    
\end{equation}
From the solutions of this cubic equation we identify the one corresponding to the $z$-coordinate of the saddle point to be 
\begin{equation}
Z_{\rm sd} = \frac{\left(\frac{2}{3}\right)^{1/3}}{h(\Delta\tau)}  + \frac{h(\Delta\tau)}{2^{1/3}\enspace3^{2/3}}, \enspace\text{where}\enspace h(\Delta\tau) = \left({\rm sign}(\Delta\tau) 9A + \sqrt{3}\sqrt{27A^2 - 4} \right)^{1/3},    
\end{equation}
with the factor $A = \left(\frac{1-s}{s}\right)\overline{\tau}$. From the above relation between the $x$- and $z$-coordinates we see that $X_{\rm sd} = \pm\sqrt{1 - Z_{\rm sd}^2}$, then we can construct the tangent map and evaluate it in the saddle point. After doing so, we find that the exponent governing the exponential growth is given by 
\begin{equation}
\lambda_{\rm saddle}^{(4,2)}(\Delta\tau) = s(\tau^*_{4,2} + \Delta\tau)\sqrt{2\left(\frac{1-s}{s}\right)^2\left(\frac{\Delta\tau}{\tau^*_{4,2} + \Delta\tau} \right)^2 - X_{\rm sd}^6},    
\end{equation}
where again we have include the appropriate prefactor which accounts for the definition of the effective Hamiltonian.

\subsubsection{The case of arbitrary even $p$}
Now that we have presented in detail the calculation of the exponents for the two models where explicit expressions can be obtained, let us briefly show some further insights which can be extracted from this approach, when we consider the cases of an arbitrary even $p$ at the $\tau^*_{p,2}$ structural instability region.

Let us start with the equations of motion for the classical flow
\begin{subequations}
\begin{align}
    \frac{dX}{dt} &= -(1-s)\overline{\tau} Y, \\
    \frac{dY}{dt} &= (1-s)\overline{\tau}X - s X^{p-1} Z, \\
    \frac{dZ}{dt} &= s X^{p-1} Y,
\end{align}
\end{subequations}
where now we have $\overline{\tau} = \frac{\delta\tau}{\tau^*_{p,2} + \Delta\tau}$. This set of equations has two fixed points at the poles, $X = Y = 0$ and $Z = \pm 1$. The other fixed points satisfy $Y = 0$ and the $x$- and $z$-coordinates are related via 
\begin{equation}
    Z = \frac{(1-s)\overline{\tau}}{sX^{p-2}}, \enspace\text{and}\enspace Z^2 = 1-X^2.
\end{equation}
These two equations give rise to a algebraic equations of degree $2p-2$ for both coordinates. For instance, the one for the $x$-coordinate si given by 
\begin{equation}
   X^{2p-2} - X^{2p-4} + \left(\frac{1-s}{s}\right)^2 \overline{\tau}^2 = 0. 
\end{equation}
Although a general solution to this algebraic equations is not available, we can use it to estimate the size of the structural instability region. Let us define the function $S(X) = X^{2p-2} - X^{2p-4} + \left(\frac{1-s}{s}\right)^2 \overline{\tau}^2$, this function has extreme points at $X = 0$ and $X_{\pm} = \pm \sqrt{\frac{2p-4}{2p-2}}$, additionally we notice that $S(1) = S(0) = \left(\frac{1-s}{s}\right)^2 \overline{\tau}^2 > 0$ are always positive. Then, the function $S(X)$ will have at least one root in the interval $X\in[0,1]$ if there is a minimum in this interval and the function evaluated at this minimum is negative. 

It is not difficult to see that $X_+$ is in fact a minimum of $S(X)$. Thus, we want conditions on the parameters such that the inequality $S(X_+) < 0$ is satisfied. Solving for those conditions we find 
\begin{equation}
\Delta\tau \le \frac{s\tau^*_{p,2}\mathcal{F}(p)}{1-s-s\mathcal{F}(p)}, \enspace\text{with}\enspace \mathcal{F}(p) = \sqrt{\frac{(p-1)(p-2)^{p-2} - (p-2)^{p-1}}{(p-1)^{p-1}}}.
\end{equation}
Thus we see that the size of the structural instability region around $\tau^*_{p,2}$ decreases with increasing $p$, with the model with $p=2$ having the most prominent one.

\subsection{Structural instability around $\tau^*_{4,4}$}
We move now to consider the structural instability region around $\tau^*_{p,4}$, and in particular will focus on the case of the model with $p=4$. Inside this instability region the effective Hamiltonian is given by 
\begin{equation}
    H_{\rm eff}^{(4,4)} = -(1-s)\overline{\tau} J_z - \frac{s}{8J^3}\left( J_x^4 + J_y^4 \right),
\end{equation}
where $\overline{\tau} = \frac{\Delta\tau}{\tau^*_{4,4} + \Delta\tau}$. The equations of motion for the asociated classical flow are 
\begin{subequations}
\begin{align}
    \frac{dX}{dt} &= \frac{s}{2}Y^3 Z - (1-s)\overline{\tau} Y, \\
    \frac{dY}{dt} &= (1-s) \overline{\tau} X - \frac{s}{2}X^3 Z, \\
    \frac{dZ}{dt} &= \frac{s}{2}X Y \left(X^2 - Y^2 \right).
\end{align}
\end{subequations}
This set of equations has fixed points on the poles, $X = Y = 0$ and $Z = \pm1$. Other fixed points can be found as solutions to $\frac{d\bm{X}}{dt}=0$, from the third of those equations we pick the condition $X\ne0$ and $Y\ne0$, leading to $X^2 = Y^2$, by substituting that condition into the first equation of the set we obtain the relation $Y^2 = \frac{2(1-s)\overline{\tau}}{sZ}$, and using this relation we can write the following cubic equation for the $z$-coordinate of the fixed point 
\begin{equation}
Z^3 - Z + \left(\frac{1-s}{s}\right)\overline{\tau} = 0,
\end{equation}
the solution of this equations which corresponds to the saddle point is given by 
\begin{equation}
Z_{\rm sd} = \frac{1 - i\sqrt{3}}{2\enspace 3^{1/3}\left( -18 A + \sqrt{3}\sqrt{-1 + 108 A^2} \right)^{1/3}} - \frac{\left( 1 + i\sqrt{3}\right)\left( -18 A + \sqrt{3} \sqrt{-1 + 108 A^2} \right)^{1/3}}{2\enspace 3^{2/3}},    
\end{equation}
where $A = \left( \frac{1-s}{s}\right)\left(\frac{\Delta\tau}{\tau^*_{4,4} + \Delta\tau} \right)$. Using this last expression we can compute the exponent of the saddle point as the largest eigenvalue of 
\begin{equation}
\mathbf{M}^{(4,4)} = \begin{pmatrix}
    0 & 2(1-s)\frac{\Delta\tau}{\tau^*_{4,4} + \Delta\tau} & \frac{s}{2}Y_{\rm sd}^3 \\
    -2(1-s)\frac{\Delta\tau}{\tau^*_{4,4} + \Delta\tau} & 0 & -\frac{s}{2}X_{\rm sd}^3 \\
    s Y_{\rm sd}^3 & -s X_{\rm sd}^3 & 0
\end{pmatrix},
\end{equation}
where $X_{\rm sd} = \pm \frac{\sqrt{1-Z_{\rm sd}^2}}{\sqrt{2}}$.

In the case of a system with arbitrary even $p\ge 4$ at this same instability, although we cannot explicitly compute the exponent, we can estimate the width of the structural instability region.

The effective Hamiltonian is 
\begin{equation}
    H_{\rm eff}^{(p,4)} = -(1-s)\overline{\tau} J_z - \frac{s}{2pJ^{p-1}}\left( J_x^p + J_y^p \right),
\end{equation}
the equations of motion of the associated classical flow are 
\begin{subequations}
\begin{align}
    \frac{dX}{dt} &= \frac{s}{2}Y^{p-1} Z - (1-s)\overline{\tau} Y, \\
    \frac{dY}{dt} &= (1-s) \overline{\tau} X - \frac{s}{2}X^{p-1} Z, \\
    \frac{dZ}{dt} &= \frac{s}{2}X Y \left(X^{p-2} - Y^{p-2} \right).
\end{align}
\end{subequations}
It is not difficult to see that the saddles are at $X\ne0$, $Y\ne0$ and they satisfy $X^{p-2} = Y^{p-2}$, furthermore $Z = \frac{2(1-s)\overline{\tau}}{sX^{p-2}}$. Using these conditions we can derive an algebraic equation for the $x$-xoordinate of the saddle poistion, it reads
\begin{equation}
    2X^{2p-2} - X^{2p-4} + 4\left( \frac{1-s}{s} \right)^2\overline{\tau}^2 = 0,
\end{equation}
to investigate the width of the structural instability region, we look for the range of parameters such that the above equations has nontrivial real solutions. To do this we consider the function $G(X) = 2X^{2p-2} - X^{2p-4} + 4\left( \frac{1-s}{s} \right)^2\overline{\tau}^2$, the extreme points of this function are at $X = 0$ and $X_{\pm} = \sqrt{\frac{p-2}{2(p-1)}}$. since $G(0)$ and $G(1)$ are positive, then if there is a minimum in the range $X\in[0,1]$ such that $G(X)$ at this minimum is negative, then the algebraic equation has at least one nontrivial solution. In fact $X_+$ is a minimum, thus we want conditions on the function parameters such that the inequality $G(X_+) < 0$ is true. After solving for those conditions we find 
\begin{equation}
    \Delta\tau \le \frac{s\tau^*_{p,4}\mathcal{G}(p)}{1 - s - s\mathcal{G}(p)},\enspace\text{with} \enspace \mathcal{G}(p) = \sqrt{\frac{(p-1)(p-2)^{p-2} - (p-2)^{p-1}}{2^p(p-1)^{p-1}}} = \frac{1}{2^p}\mathcal{F}(p),
\end{equation}
thus this region of structural instability shrinks with increasing $p$ and is exponentially narrower that the region of structural instability around $\tau^*_{p,2}$.

\end{widetext}
\newpage

\bibliography{p_spin}
\end{document}